\documentclass[reprint,superscriptaddress,amsmath,amssymb,aps,prl,longbibliography]{revtex4-2}

\usepackage{graphicx,tabularx,makecell}
\usepackage{amsmath,amsfonts,amssymb}
\usepackage{mathtools}
\usepackage{bm,dsfont}
\usepackage[dvipsnames]{xcolor}
\usepackage[bookmarks=true,colorlinks=true,urlcolor=blue,linkcolor=blue,citecolor=blue,breaklinks]{hyperref}
\usepackage[all]{hypcap}
\usepackage{ulem}

\graphicspath{{figures/}}

\renewcommand{\arraystretch}{1.3}
\definecolor{pu}{RGB}{200,50,200}
\definecolor{gr}{RGB}{0,187,0}
\definecolor{bl}{RGB}{68,34,200}
\definecolor{re}{RGB}{200,34,68}
\definecolor{ye}{RGB}{255,165,0}
\definecolor{oran}{RGB}{255,170,0}

\newcommand{\RNum}[1]{\uppercase\expandafter{\romannumeral #1\relax}}

\begin{document}

\title{Role of electron-electron interactions in $M$-valley twisted transition metal dichalcogenides}

\author{Christophe De Beule}
\affiliation{Department of Physics and Astronomy, University of Pennsylvania, Philadelphia, Pennsylvania 19104, USA}
\affiliation{Department of Physics, University of Antwerp, Groenenborgerlaan 171, 2020 Antwerp, Belgium}
\author{Liangtao Peng}
\affiliation{Department of Physics, Washington University in St. Louis, St. Louis, Missouri 63130, United States}
\author{E. J. Mele}
\affiliation{Department of Physics and Astronomy, University of Pennsylvania, Philadelphia, Pennsylvania 19104, USA}
\author{Shaffique Adam}
\affiliation{Department of Physics, Washington University in St. Louis, St. Louis, Missouri 63130, United States}
\affiliation{Department of Physics and Astronomy, University of Pennsylvania, Philadelphia, Pennsylvania 19104, USA}
\affiliation{Department of Materials Science and Engineering, 
National University of Singapore, 9 Engineering Drive 1, 
Singapore 117575}

\date{\today}

\begin{abstract}
We investigate the role of long-range Coulomb interactions in $M$-valley moir\'es using the self-consistent Hartree-Fock approximation. This platform was recently proposed [\href{https://www.nature.com/articles/s41586-025-09187-5}{Nature \textbf{643}, 376 (2025)} and \href{https://arxiv.org/abs/2411.18828}{arXiv:2411.18828 (2024)}] as a new class of experimentally realizable moir\'e materials using twisted transition metal dichalcogenides homobilayers with the 1T structure.
While these seminal studies considered the noninteracting theory without an electric displacement field, this work shows that both electron-electron interactions at finite doping and an interlayer bias strongly modify the moir\'e bands. 
For small twists ($\lesssim 5^\circ$) the density of states versus filling and interlayer bias displays qualitatively different behavior for twisting near aligned ($0^\circ$) and antialigned ($60^\circ$) stacking with tunable Van Hove singularities (VHSs). Moreover, interactions pin the VHS to the Fermi energy over a finite range of doping both at zero and finite bias depending on the stacking type, an effect known to enhance both superconductivity and strongly correlated states. At half filling, we obtain the phase diagram as a function of interaction strength, interlayer bias, and twist angle. We find a competition driven by band mixing between an isotropic ferromagnet and an antiferromagnet that are nearly degenerate over a wide range of experimentally accessible parameters.
Our work demonstrates that correlated states in $M$-valley 1T tTMDs can be strongly tuned \textit{in situ} both by applying an electric displacement field and by electron doping.
\end{abstract}

\maketitle

Correlated states in moir\'e materials that emerge by partially filling energetically flattened bands have been intensively explored in hexagonal systems where the low-energy degrees of freedom originate from high-symmetry points such as the $K/K'$ and $\Gamma$ points of the Brillouin zone \cite{lopesdossantos_graphene_2007,suarez_morell_flat_2010,bistritzer_moire_2011,wu_topological_2019,angeli_gamma_2021}. A prominent example is magic-angle twisted bilayer graphene where the bands near charge neutrality arise from Dirac cones at the $K/K'$ valleys, leading to an approximate $\mathrm{U}(4) \times \mathrm{U}(4)$ symmetry in the low-energy sector combining spin, valley, and sublattice degrees of freedom \cite{bultinck_ground_2020}. This system hosts a range of interaction-driven phenomena, including correlated insulators \cite{cao_correlated_2018}, superconductivity \cite{cao_unconventional_2018}, and intervalley coherence \cite{nuckolls_quantum_2023}. Another prominent example are twisted transition metal dichalcogenides (TMDs) such as twisted bilayer 2H WSe$_2$ and MoTe$_2$ where strong Ising spin-orbit coupling locks the spin to valley at the $K/K'$ point in the valence band. Here interactions give rise to superconductivity \cite{xia_superconductivity_2025,guo_superconductivity_2025,xu_signatures_2025}, itinerant magnetism \cite{knuppel_correlated_2025,ghiotto_stoner_2024}, and the fractional anomalous quantum Hall effect \cite{zeng_thermodynamic_2023,park_observation_2023}. A parallel line of work has focused on $\Gamma$-valley moir\'e TMDs, where the low-energy bands are valley-less. Examples include the valence band of twisted homobilayers of 2H MoS$_2$ or 2H MoSe$_2$ \cite{angeli_gamma_2021} which display correlated states such as Wigner crystals \cite{kaushal_magnetic_2022}, enabled by the flatness of the $\Gamma$-valley moir\'e bands.

It is also possible to have twisted materials with low-energy valleys located at the three inequivalent M points of the hexagonal Brillouin zone 
\cite{kariyado_flat_2019,fujimoto_perfect_2022,kariyado_twisted_2023,bao_anisotropic_2025}.  Very recently two theoretical groups independently proposed that twisting experimentally exfoliable materials such as SnSe$_2$ and group 4 TMDs such as ZrS$_2$ or HfS$_2$ result in $M$-valley moir\'es \cite{lei_moire_2024,calugaru_moire_2025}. These materials are insulators down to the monolayer with the hexagonal 1T structure and point group $D_{3d}$, and their conduction band minima generically occurs at the M points \cite{zhao_elastic_2017}.
The resulting moir\'e bands carry a threefold valley degree of freedom and have an approximate spin degeneracy inherited from the monolayer. For sufficiently small twist angles, the dispersion of the lowest moir\'e bands become negligible, giving rise to an effective $\mathrm{SU}(3)$ valley symmetry, which, when combined with spin, yields an emergent $\mathrm{SU}(6)$ flavor symmetry.

In this paper, we study the role of electron-electron interactions in the bottommost moir\'e bands of $M$-valley moir\'e systems in the weak-coupling regime, using both aligned and antialigned 1T tSnSe$_2$ as representative examples. To this end, we use the self-consistent Hartree-Fock method applied to the moir\'e continuum model that fully takes into account band mixing. In the absence of symmetry breaking, we find that the moir\'e bands are strongly renormalized by interactions as the electron filling increases from charge neutrality ($\nu = 0$) to full filling $(\nu = 6)$ of the low-energy moir\'e band manifold. In particular, at half filling we find that the bandwidth increases significantly. While this would ordinarily disfavor correlated states, we also find surprisingly that the Van Hove singularity in the density of states gets pinned to the Fermi energy in the presence of interactions which enhances the likelihood of strongly correlated states and superconductivity as observed in other moir\'e systems such as twisted bilayer graphene near the magic angle.  

Mapping the density of states as a function of filling and applied interlayer bias shows a qualitatively different behavior for small twist angles near aligned ($0^\circ$) and antialigned ($60^\circ$) stacking. Both cases yield a moir\'e that varies slowly with respect to the atomic scale, but are inequivalent due to the lack of $180^\circ$ in-plane rotation symmetry of the 1T monolayer. We then study symmetry-broken phases at half filling that spontaneously break time-reversal symmetry and/or rotation symmetry by flavor polarization, but preserve moir\'e translational symmetry and valley charge conservation.
Our study establishes a baseline for correlation effects in $M$-valley systems, providing a framework for interpreting ongoing and future experimental results on $M$-valley moir\'e TMDs.
\begin{figure}
    \centering
    \includegraphics[trim={0cm 0 0cm 0},clip,width=\linewidth]{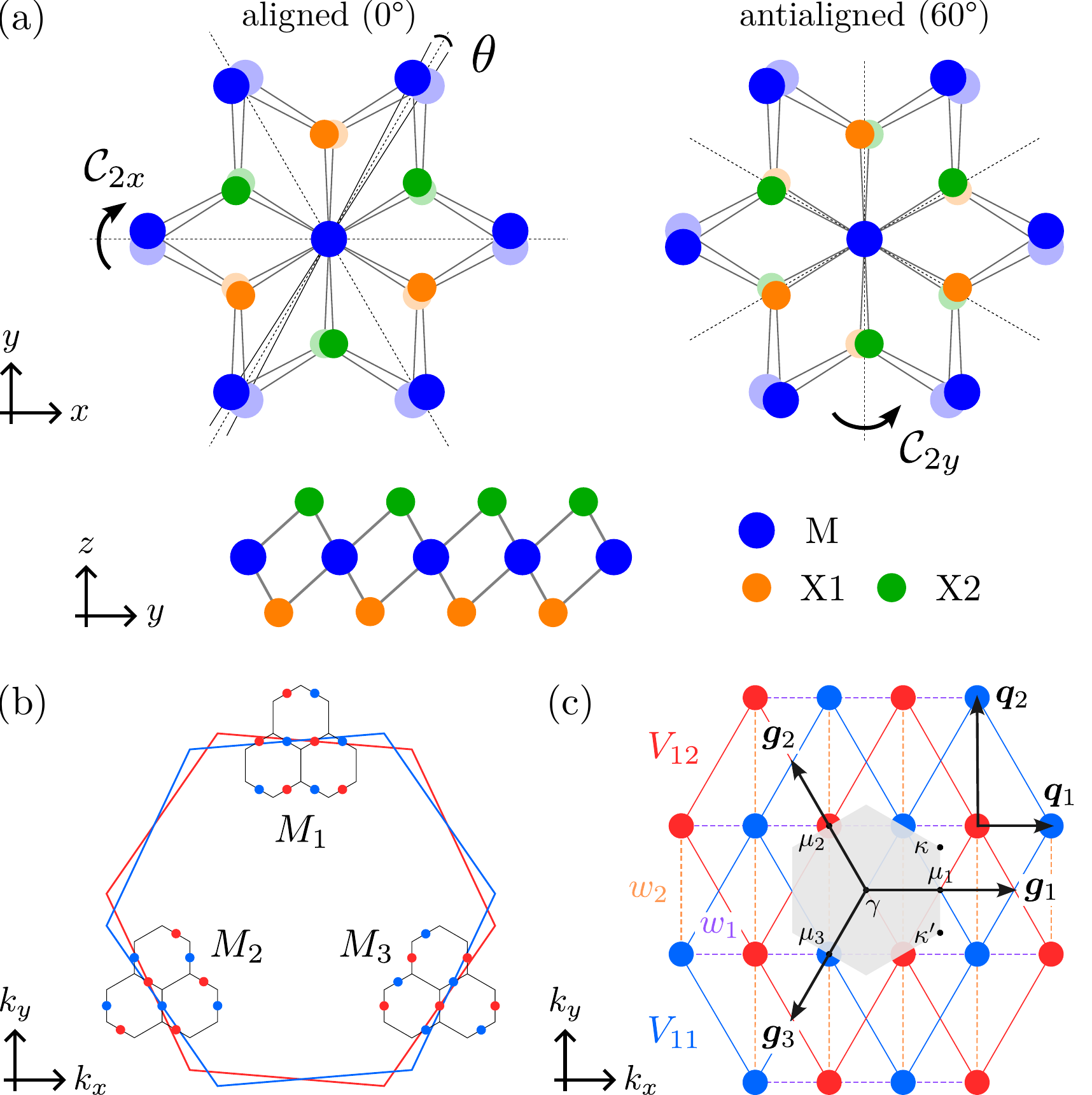}
    \caption{Real space and momentum space representations for $M$-valley moir\'e materials. (a) Lattice structure of twisted aligned and antialigned homobilayer 1T TMDs MX$_2$. Here X1 (X2) is buckled into (out of) the plane, see bottom panel, and the dark and light atoms correspond to layer $l=1,2$ which is rotated by $\pm \theta/2$, respectively. Shown for $\theta = 5.09^\circ$ close to the twist center which is chosen here as eclipsing metal atoms. For a general twist center the $D_3 = \left< \mathcal C_{3z}, \mathcal C_{2x} \right>$ symmetry is an approximate emergent symmetry on the moir\'e scale for $L/a \gg 1$. (b) Brillouin zone of the twisted monolayers (red and blue large hexagons) showing the three $M$ valleys (dots) for both layers and the moir\'e Brillouin zone (small hexagons). (c) Illustration of the moir\'e potentials in momentum space for the $M_1$ valley ($\tau = 1$). Solid (dashed) lines represent intralayer (interlayer) processes and the gray hexagon is the MBZ. Intralayer terms along $\pm \bm g_1$ are not shown for clarity.}
    \label{fig:fig1}
\end{figure}

\textcolor{NavyBlue}{\textit{Model}} --- We consider a twisted homobilayer made from $M$-valley monolayers with point group $D_{3d}$. Here layers $l=1,2$ are rotated by an angle $\pm \theta/2$, respectively, with respect to aligned ($0^\circ$) or antialigned ($60^\circ$) stacking, see Fig.\ \ref{fig:fig1}(a). In the long-wavelength limit, i.e.\ for small twist angles, the emergent moir\'e lattice has period $L = a/\theta \gg a$ with $a$ the monolayer lattice constant, and emergent point group $D_3$ generated by $\mathcal C_{3z}$ and $\mathcal C_{2x}$ ($\mathcal C_{2y}$) for twisting near aligned (antialigned) stacking in the coordinate system of Fig.\ \ref{fig:fig1}(a).

In particular, we consider semiconductors such as 1T SnSe$_2$ or 1T ZrS$_2$ with three spin degenerate conduction band minima at the three $M_\tau$ points ($\tau = 1,2,3$) \cite{zhao_elastic_2017}, illustrated in Fig.\ \ref{fig:fig1}(b). The corresponding low-energy moir\'e theory for the conduction band can thus be written as $H = H_0 + H_\mathrm{ee}$ where $H_0 = \sum_{s=\uparrow,\downarrow} \sum_{\tau=1}^3  H_{s\tau}$ is the single-particle Hamiltonian and $H_\mathrm{ee}$ is a gate-screened Coulomb interaction. While the twisted bilayer lacks inversion symmetry, an approximate SU(2) symmetry is inherited from the monolayer \cite{lei_moire_2024,calugaru_moire_2025}. Therefore we can write $H_{s\tau} = \int d^2 \bm r \,  \psi_{s\tau}^\dag \mathcal H_\tau  \psi_{s\tau}$ with
\begin{equation} \label{eq:Htau}
    \mathcal H_\tau = \begin{bmatrix} \left( \tfrac{-\hbar^2}{2m_\tau} \right)_{ij} \partial^i \partial^j + V_{\tau1} & T_\tau^* \\ T_\tau & \left( \tfrac{-\hbar^2}{2m_\tau} \right)_{ij} \partial^i \partial^j + V_{\tau2} \end{bmatrix},
\end{equation}
where summation over repeated indices is implied, and $\psi_{s\tau} = ( \psi_{s\tau1}, \psi_{s\tau2} )^\top$ are fermion field operators in layer space that satisfy $\{ \psi_{s\tau l}^\dag(\bm r), \psi_{s'\tau'l'}(\bm r') \} = \delta_{ss'} \delta_{\tau\tau'} \delta_{ll'} \delta(\bm r-\bm r')$. The M points are time-reversal invariant momenta that are related by $\mathcal C_{3z}$ rotations such that $H_{s,\tau+1} = \mathcal C_{3z} H_{s\tau} \mathcal C_{3z}^{-1}$ with $\mathcal C_{3z} \psi_{s\tau}(\bm r) \mathcal C_{3z}^{-1} = \psi_{s,\tau+1}(\mathcal C_{3z}\bm r)$. Hence the Hamiltonian for each $\tau$ is only constrained by time-reversal symmetry, moir\'e translations, and e.g.\ for $\tau = 1$ by $\mathcal C_{2x}$ ($\mathcal C_{2y}$) for aligned (antialigned) stacking where we take $M_1 = 2\pi \hat y / (\sqrt{3}a)$ for the untwisted bilayer. We thus only specify $\mathcal H_1$ in the following. The monolayer $\mathcal C_{2x}$ symmetry implies an inverse effective mass tensor $1/m_1 =\text{diag}(m_\perp^{-1}, m_\parallel^{-1})$ and in lowest order, the moir\'e potentials can be written as
\begin{align}
    V_{1l}(\bm r) & = (-1)^{l+1} \frac{V_z}{2} + \sum_{n=1}^3 2v_{ln} \cos(\bm g_n \cdot \bm r + \psi_{ln}), \label{eq:V1l}\\
    T_1(\bm r) & = T_1^*(\bm r) = \sum_{n=1}^2 2w_n \cos(\bm q_n \cdot \bm r + \phi_n), \label{eq:T1}
\end{align}
where $V_z$ is an interlayer bias due to an external electric field perpendicular to the layers, which breaks the out-of-plane twofold rotation symmetry. Here the $\bm g_n$ are the shortest nonzero moir\'e reciprocal vectors related by $\mathcal C_{3z}$ with $\bm g_1 = 4\pi \hat x/(\sqrt{3}L)$, $\bm q_1 = \bm g_1/2$, and $\bm q_2 = \bm q_1 + \bm g_2$. These correspond to the dominant intralayer and interlayer moir\'e scattering processes, respectively. We can understand this intuitively from the geometry of the moir\'e Brillouin zone (MBZ) in a given $M$ valley, as shown in Fig.\ \ref{fig:fig1}(c). Twofold rotation symmetry further requires that $v_{21} = v_{11}$, $v_{22} = v_{13}$, and $v_{23} = v_{12}$. In addition, $\psi_{21} = \pm \psi_{11}$, $\psi_{22} = \pm \psi_{13}$, and $\psi_{23} = \pm \psi_{12}$, for aligned ($\mathcal C_{2x}$) and antialigned ($\mathcal C_{2y}$) stacking, respectively. For the interlayer tunneling, we further have $\phi_2 = 0$ for aligned stacking and $\phi_1 = 0$ for antialigned stacking. See Supplemental Material (SM) \footnote{See Supplemental Material at [insert link] for a derivation of the moir\'e continuum Hamiltonian from symmetry} for details. 

Furthermore, in the local-stacking approximation \cite{jung_origin_2015}, the inversion ($\bm r \mapsto -\bm r$ and $z \mapsto -z$) and mirror ($z \mapsto -z$) symmetry of untwisted aligned and antialigned bilayers implies $V_{\tau1}(\bm r) = V_{\tau2}(\pm \bm r)$, respectively  \cite{lei_moire_2024,calugaru_moire_2025}. In this case, the theory for $V_z = 0$ has additional intravalley model symmetries, given respectively by mirror $\mathcal M_z \psi_{s\tau}(\bm r) \mathcal M_z^{-1} = \sigma_x \psi_{s\tau}(\bm r)$ and inversion $\mathcal P \psi_{s\tau}(\bm r) \mathcal P^{-1} = \sigma_x \psi_{s\tau}(-\bm r)$. The latter results in $T_\tau(\bm r) = T_\tau(-\bm r)$ for antialigned stacking. Incidentally, we notice a previously unreported duality between these approximate moir\'e symmetries inherited from untwisted bilayers, for aligned (antialigned) 1T tTMDs such as tSnSe$_2$ and tZrS$_2$ with monolayer point group $D_{3d}$, and antialigned (aligned) 2H tTMDs such as tWSe$_2$ and tMoTe$_2$ with monolayer point group $D_{3h}$. This immediately implies that antialigned 1T tTMDs contain two degenerate low-energy stacking configurations in the moir\'e cell and relax towards a triangular soliton network of partial shear dislocations, while aligned 1T tMDS only have one low-energy configuration resulting in a honeycomb network of full shear dislocations \cite{de_beule_theory_2025}.

We note that for the choice of basis used to write Eq.\ \eqref{eq:Htau} the tunneling is not moir\'e periodic: $T_1(\bm r + \bm L) = (-1)^m T_1(\bm r)$ with $\bm g_1 \cdot \bm L = 2\pi m$ and integer $m$. The advantage of this basis is that symmetries act locally in momentum space, because we gauged the momentum of the rotated $M_\tau$ points of layer $l$ denoted as $\bm \mu_{\tau l}$. On the other hand, because $\mathcal H_\tau$ is only quasiperiodic in this basis, it obfuscates the fact that for aligned stacking, the local-stacking symmetry $\mathcal M_z$ is nonsymmorphic in momentum space because it exchanges $\bm \mu_{\tau 1}$ and $\bm \mu_{\tau 2}$ which are separated by \textit{half} a reciprocal lattice vector \cite{calugaru_moire_2025}. Consequently, $\mathcal M_z$ does not commute with certain moir\'e translations, e.g.\ $\bm L_1$ for $\tau = 1$ with $\bm g_1 \cdot \bm L_1 = 2\pi$ \cite{calugaru_moire_2025}. Hence, for each valley the $\mathcal M_z$ eigenbasis realizes a projective representation of the translation group equivalent to threading half a flux quantum (which is time-reversal invariant). In the $\mathcal M_z$ eigenbasis, the moir\'e cell is doubled yielding two orbitals with opposite mirror eigenvalues for the low-energy mor\'e bands \cite{calugaru_moire_2025}. 

To diagonalize $H_0$ by Fourier transform, we restore the origin by letting $\psi_{s\tau l} \rightarrow e^{-i \bm \mu_{\tau l} \cdot \bm r} \psi_{s\tau l}$ 
such that $T_1(\bm r) \rightarrow e^{-i\bm g_1 \cdot \bm r / 2} T_1(\bm r)$. This yields Bloch eigenstates $| \phi_{\tau n\bm k}^{(l)} \rangle = \sum_{\bm g} u_{\tau n\bm k}^{(l)}(\bm g) \left| \bm k+\bm g \right>$ with energies $\varepsilon_{\tau n\bm k} = \varepsilon_{\tau+1,n,\mathcal C_{3z}\bm k}$ where $\tau$ is defined modulo $3$. The model in Eq.\ \eqref{eq:Htau} and the moir\'e potentials from Eqs.\ \eqref{eq:V1l} and \eqref{eq:T1} were previously constructed and studied in Refs.\ \cite{lei_moire_2024,calugaru_moire_2025} at charge neutrality in the absence of an interlayer bias $V_z$. In what follows we use parameters for tSnSe$_2$ fitted to density-functional theory (DFT) \cite{calugaru_moire_2025} and which are listed in the SM \cite{Note1}. Here we have chosen parameters that reproduce well the low-energy moir\'e bands at charge neutrality obtained from DFT. For aligned stacking, these parameters obey the local-stacking approximation, while for antialigned stacking, this is only approximately true.
\begin{figure}
    \centering
    \includegraphics[trim={0cm 0 0cm 0},clip,width=\linewidth]{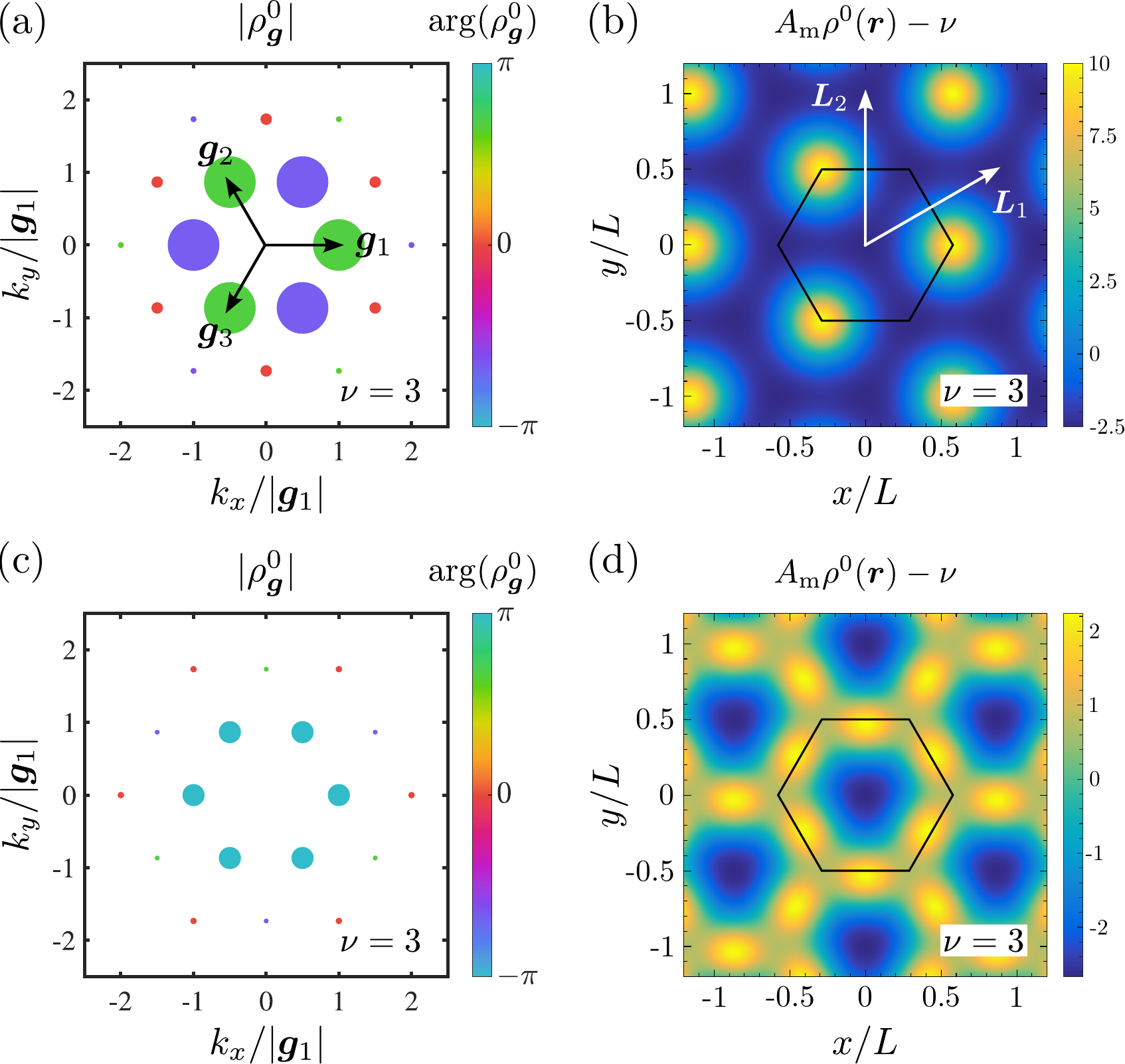}
    \caption{Large moir\'e scale density fluctuations in the noninteracting model highlight the importance of Coulomb interactions. (a) Density fluctuations for aligned $3.89^\circ$ 1T tSnSe$_2$ at half filling of the low-energy moir\'e bands ($\nu = 3$) in the absence of electron-electron interactions. The size gives the magnitude $|\rho_{\bm g}^0|$ and the color its phase $\arg(\rho_{\bm g}^0)$. (b) Corresponding density $\rho^0(\bm r)$ showing only the fluctuations in units of $A_\text{m}^{-1} = 2 / (\sqrt{3}L^2)$, and where the hexagon is the moir\'e cell. The large amplitude on the moir\'e scale gives a strong Hartree potential. The density respects $D_3$ because we sum over all three $M$ valleys. (c) Density fluctuations for antialigned stacking using the same twist angle and scale as in (a). (d) Same as (b) for antialigned stacking. In this case, the density fluctuations are much smaller as compared to aligned stacking for the same twist, giving a weaker Hartree effect. Because the density for each $M$ valley is related by $\mathcal C_{3z}$ symmetry, they roughly coincide for aligned stacking, forming a triangular lattice with three orbitals at each site \cite{calugaru_moire_2025}. On the other hand, for antialigned stacking, we see that each of the three valleys is localized on one sublattice of a kagome lattice.}
    \label{fig:fig2}
\end{figure}

\textcolor{NavyBlue}{\textit{Interaction-renormalized moir\'e bands}} --- We now investigate the effect of electron-electron interactions on the low-energy $M$-valley moir\'e bands using the self-consistent Hartree-Fock (HF) approximation implemented in band basis \cite{peng_many-body_2025,Note1}. In particular, we consider a dual-gate screened Coulomb interaction 
$U_{\bm q} = e^2\tanh{(qd)}/(2\epsilon \epsilon_0 q)$ where $d$ is the gate-to-sample distance. Throughout this work, we set $d = 20\,\mathrm{nm}$ and treat the relative dielectric constant $\epsilon$ of the dielectric spacer as a phenomenological parameter that tunes the interaction strength.  We anticipate that $\epsilon \approx 25$ corresponds to the physically relevant scenario since this value reproduces experimental features for other TMD moir\'e systems e.g.\ 2H-tWSe$_2$ \cite{knuppel_correlated_2025,peng_magnetism_2025}.
\begin{figure}[t]
    \centering
    \includegraphics[trim={0cm 0 0cm 0},clip,width=1\columnwidth]{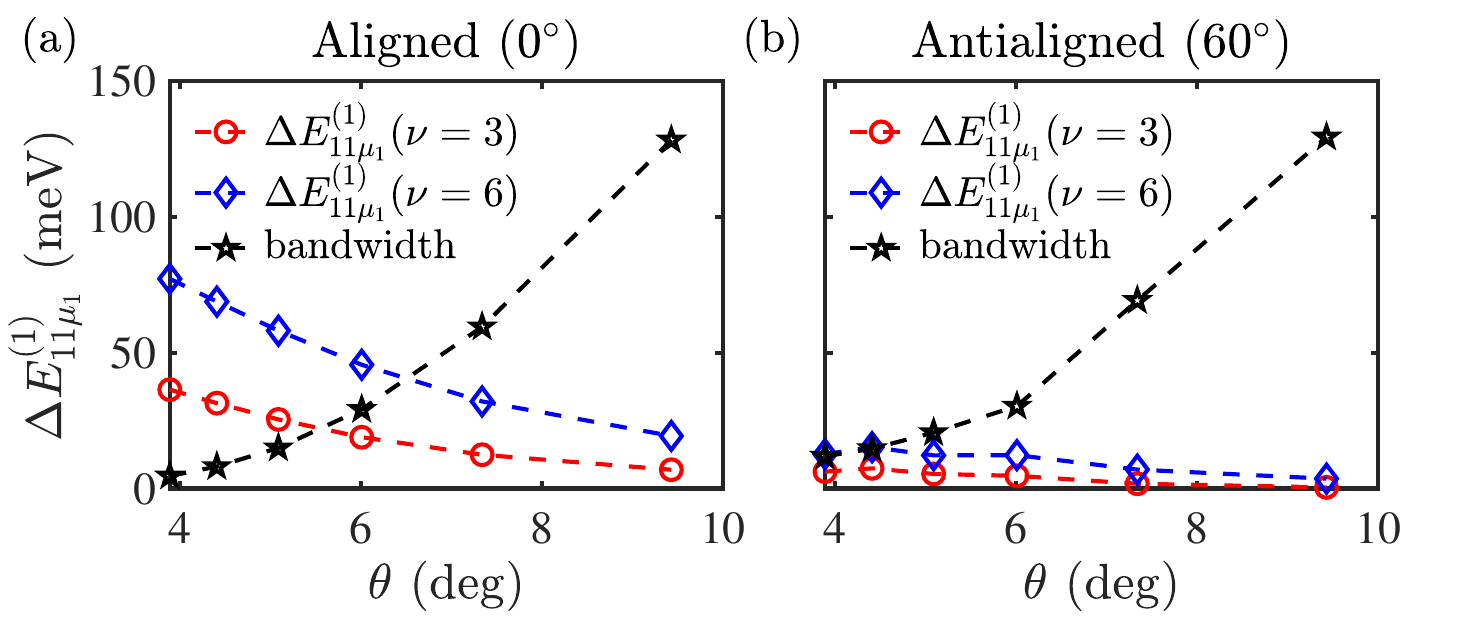}
    \caption{Comparison of one-shot Hartree correction and bandwidth as a function of twist angle for (a) aligned and (b) antialigned 1T tSnSe$_2$. Figures use $\epsilon = 25$, $V_z =0$, and valley $\tau = 1$ at the $\mu_1$ point of the MBZ for $\nu = 3$ and $\nu = 6$.}
    \label{fig:fig3}
\end{figure}

To illustrate the effect of interactions in the weak-coupling regime, we first consider the Hartree contribution written as $H_\text{H} = \sum_{s,\tau,\bm k} \sum_{\bm g\neq\bm 0} \rho_{\bm g} U_{\bm g} c_{s\tau,\bm k+\bm g}^\dag c_{s\tau,\bm k}$ where this mean-field electrostatic interaction depends on the carrier density fluctuations given by 
\begin{equation}
    \rho_{\bm g} = \frac{2}{N_\text{m} A_\text{m}} \sum_{\tau,n,\bm k \in \mathrm{occ}} \sum_{\bm g',l} \left[ u_{\tau n\bm k}^{(l)}(\bm g') \right]^* u_{\tau n\bm k}^{(l)}(\bm g + \bm g').
\end{equation}
Here the factor $2$ accounts for spin, $A_\text{m}$ the moir\'e cell area, $N_\text{m}$ the number of cells, i.e.\ the number of $\bm k$ points in our grid, $n$ is the band index, and $u_{\tau n\bm k}^{(l)}(\bm g)$ are Fourier amplitudes of the eigenstates of the mean-field Hamiltonian. We show $\rho_{\bm g}$ in the absence of interactions in Fig.\ \ref{fig:fig2}(a) for aligned 1T tSnSe$2$ with $\theta = 3.89^\circ$ at half filling of the low-energy moir\'e band manifold ($\nu = 3$). We find that the density fluctuations have significant support only on the first three reciprocal stars with the dominant contribution coming from the first star. They also obey $D_3 = \left< \mathcal C_{3z}, \mathcal C_{2x} \right>$ symmetry because all three $M$ valleys are equally filled in the noninteracting case. In particular, the first star is not constrained by $\mathcal C_{2x}$ and is generally complex. In real space, the density is strongly localized on a triangular lattice and on the order of $10A_\text{m}^{-1}$, see Fig.\ \ref{fig:fig2}(b). Because the $M$ valleys are related by $\mathcal C_{3z}$ one obtain a triangular lattice with three orbitals corresponding to the valleys \cite{calugaru_moire_2025}. On the other hand, for antialigned stacking, the density fluctuations are much weaker, as shown in Fig.\ \ref{fig:fig2}(c). Moreover, now $D_3 = \left< \mathcal C_{3z}, \mathcal C_{2y} \right>$ makes the first star real, such that the density has approximate $\mathcal C_{6z}$ symmetry, as can be seen in Fig.\ \ref{fig:fig2}(d). One obtains a kagome lattice with each $M$ valley localized on one sublattice \cite{bernevig_talk_2025}. To further illustrate the Hartree effect, we plot the one-shot Hartree shift $\Delta E_{\tau n \bm k}^{(1)} = \sum_{\bm g \neq \bm 0} \rho_{\bm g} U_{\bm g}\, \Lambda_{\tau n \bm k}(\bm g)$ of the bottom band for $\tau = 1$ at the $\mu_1$ point of the moir\'e Brillouin zone as a function of twist angle in Fig.\ \ref{fig:fig3} for both aligned and antialigned stacking. Here we define the form factor $\Lambda_{\tau n \bm k}(\bm g) = \sum_{l,\bm g'} [ u_{\tau n \bm k}^{(l)}(\bm g') ]^* u_{\tau n \bm k}^{(l)}(\bm g' + \bm g)$. Similarly to twisted bilayer graphene near the magic angle, strong density modulations on the moir\'e scale yield a significant Hartree effect (see e.g. Ref.~\cite{ezzi_self-consistent_2024}, and references therein). By contrast the Hartree potential can often be neglected for $K$ valley tTMDs such as aligned 2H-tWSe$_2$ (see e.g.\ Ref.\ \cite{peng_magnetism_2025}, and references therein).
\begin{figure}[t]
    \centering
    \includegraphics[trim={0cm 0 0cm 0},clip,width=1\columnwidth]{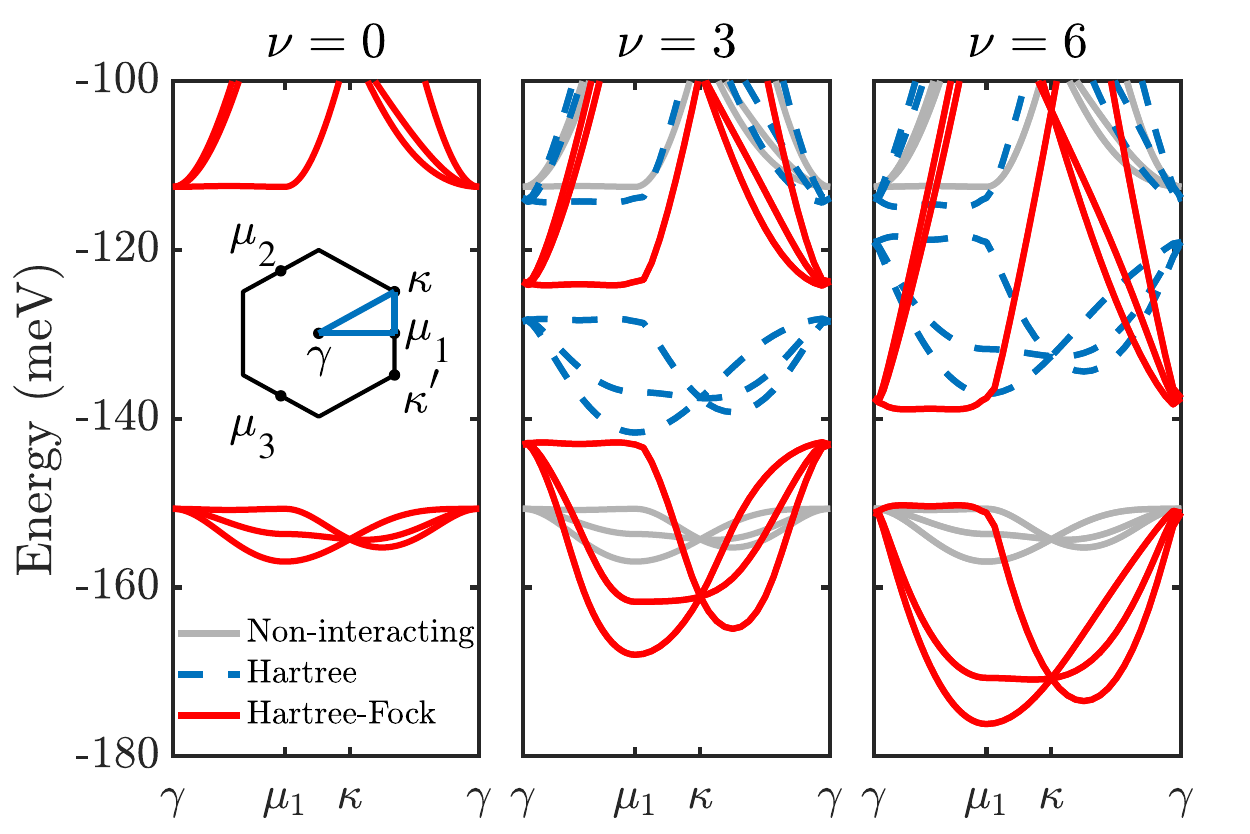}
    \caption{Hartree-Fock bands for $\epsilon = 25$ of the symmetry unbroken state for \textit{aligned} $3.89^\circ$ 1T tSnSe$_2$ with interlayer bias $V_z = 30$ meV, shown for filling factors $\nu = 0$, $3$, and $6$. A finite $V_z$ splits the bands of the three valleys which are otherwise pairwise degenerate along the chosen path. At $\nu = 0$ (left panel) the net density vanishes and the bands are those fitted to DFT, giving three low-energy bands, one for each valley \cite{calugaru_moire_2025}. As the filling increases (middle and right panels) Hartree and Fock corrections widen the low-energy band manifold and reduce the gap to the high-energy dispersive bands.}
    \label{fig:fig4}
\end{figure}
\begin{figure}[t!]
    \centering
    \includegraphics[trim={0cm 0 0cm 0},clip,width=1\columnwidth]{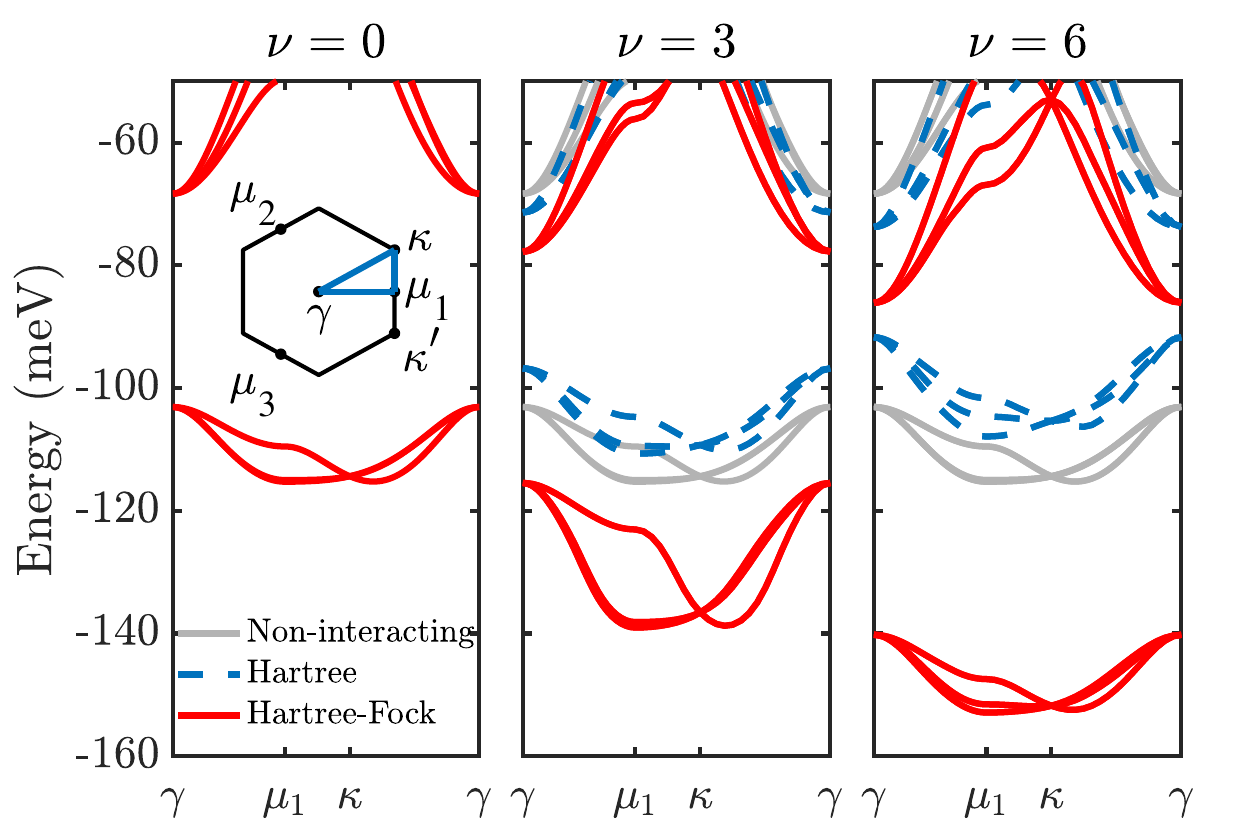}
    \caption{HF bands for $\epsilon = 25$ of the symmetry unbroken state for \textit{antialigned} $3.89^\circ$ 1T tSnSe$_2$ with $V_z = 30$ meV, shown for filling factors $\nu = 0$, $3$, and $6$. A finite $V_z$ splits the bands of the three valleys which are otherwise pairwise degenerate along the shown path. At $\nu = 0$ (left panel) the net density vanishes and the bands are those fitted to DFT, giving three low-energy bands, one for each valley \cite{calugaru_moire_2025}. As electrons are added (middle and right panels) HF corrections lead to an increase of the gap to remote bands, unlike for aligned stacking. The bandwidth initially increases ($\nu = 3$) but then decreases back for larger filling ($\nu = 6$).}
    \label{fig:fig5}
\end{figure}

Having established that $M$-valley moir\'e bands feature a significant Hartree potential, we now consider the symmetry unbroken HF state. This state serves as a \textit{parent state} from which symmetry-broken ground states may emerge. While the symmetry unbroken state is generally not the HF ground state, it is a local energy minimum of the mean field Hamiltonian. We find that both Hartree and Fock contributions play a dominant role in band renormalization. 
In Fig.\ \ref{fig:fig4}, we show the HF band structures for aligned 1T tSnSe$_2$ with $\theta = 3.89^\circ$ at filling factors $\nu = 0$, $3$, and $6$. At charge neutrality ($\nu = 0$) the bands are given by $H_0$ obtained from the fit to DFT, featuring an isolated and flattened low-energy manifold. As electrons are added, the Hartree term increases the bandwidth and reduces the gap between the lowest and remote bands, while the Fock term tends to enhance both the bandwidth and gap. These effects compete, resulting in a non-monotonic band evolution with filling. Similarly, for antialigned stacking, shown in Fig.\ \ref{fig:fig5} the HF corrections strongly modify the low-energy moir\'e bands. Though in this case, the bandwidth first increases with filling then saturates as one approaches full filling.

\begin{figure}
    \centering
    \includegraphics[width=\linewidth]{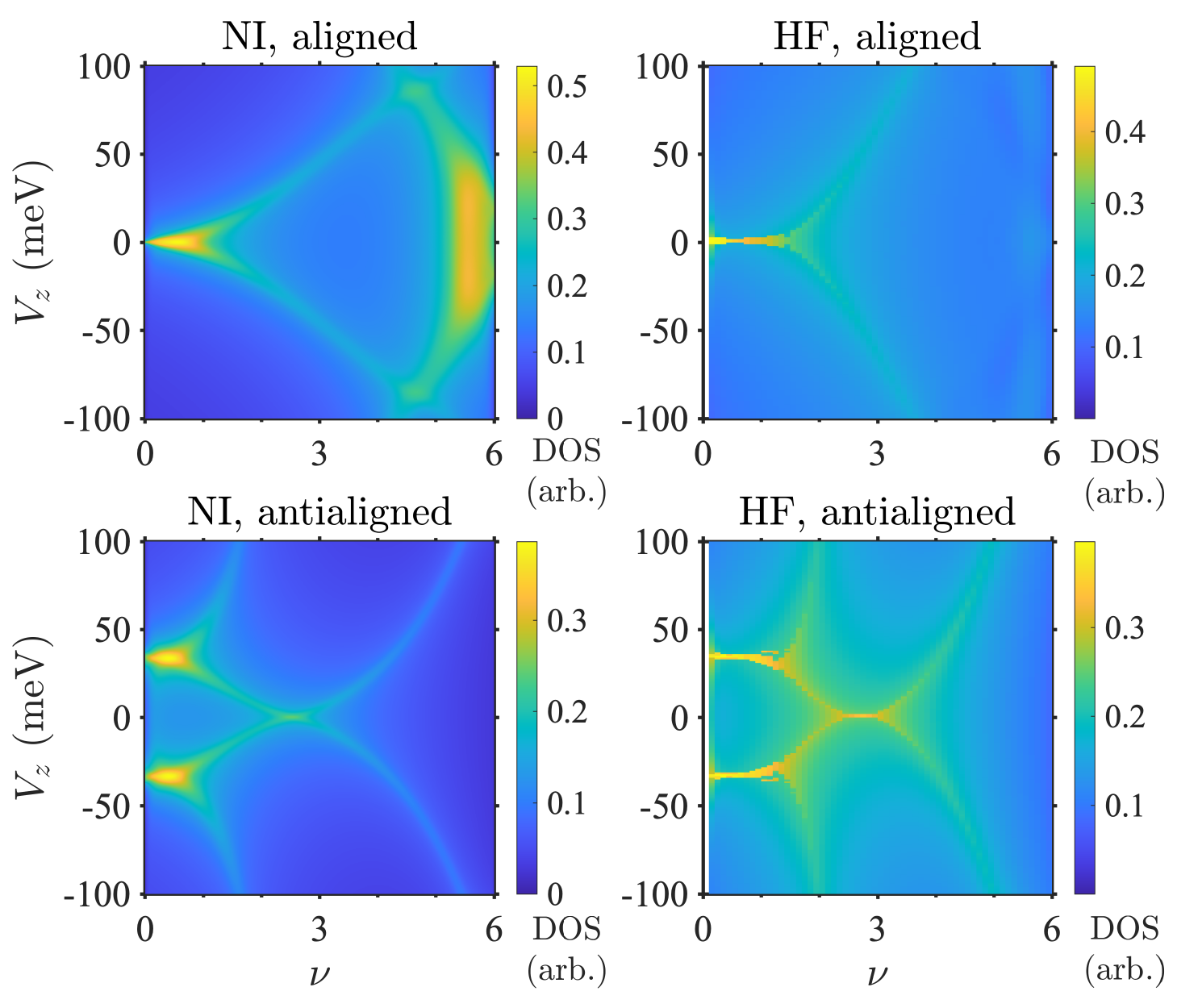}
    \caption{Density of states (DOS) of the noninteracting (left) and \textit{symmetry unbroken} Hartree-Fock state (right) as a function of filling factor $\nu$ (number of electrons per moir\'e cell) and interlayer bias $V_z$ due to an applied electric displacement field. We consider $3.89^\circ$ 1T tSnSe$_2$ for both aligned (top) and antialigned (bottom) stacking. In the noninteracting case, the DOS for $V_z=0$ shows Van Hove singularities (VHSs) near the band edges for aligned stacking and just below half filling for antialigned stacking, that disperse in the $\nu-V_z$ plane. Interestingly, for antialigned stacking the VHS moves towards charge neutrality as $|V_z|$ increases. Interactions modify this picture: HF corrections soften the Lifshitz transition and removes the VHS near full filling for aligned stacking. Moreover, the VHSs become pinned to the Fermi energy near charge neutrality, and also near half filling for antialigned stacking which give rise to the ``Y shaped'' features.} 
    \label{fig:fig6}
\end{figure}

The interaction-renormalized moir\'e bands are further illustrated in Fig.\ \ref{fig:fig6}, which shows the density of states (DOS) as a function of interlayer bias $V_z$ and filling factor $\nu$. We show the DOS for both the noninteracting (left) and HF (right) case, for both aligned (top) and antialigned (bottom) stacking, using parameters for $3.89^\circ$ tSnSe$_2$. The DOS is symmetric in $V_z$ because $V_z$ is odd under twofold out-of-plane rotation symmetry of the moir\'e. In the absence of interactions and $V_z = 0$, there is a Van Hove singularity (VHS) close to charge neutrality ($\nu = 0$) for aligned stacking due to two saddle points in the lowest moir\'e band at $k_x = 0$ and $k_y = \pm \pi/L$ for $\tau = 1$. As $|V_z|$ increases, the saddle points merge with a minimum, giving rise to a high-order VHS \cite{yuan_magic_2019} at a critical value of $V_z$. The resulting saddle point then becomes pinned at $\bm \mu_{\tau1}$ or $\bm \mu_{\tau2}$ depending on the sign of $V_z$. Another VHS exists close to full filling where the band is flattened ($\nu = 6$).

\begin{figure*}
    \centering
    \includegraphics[width=\linewidth]{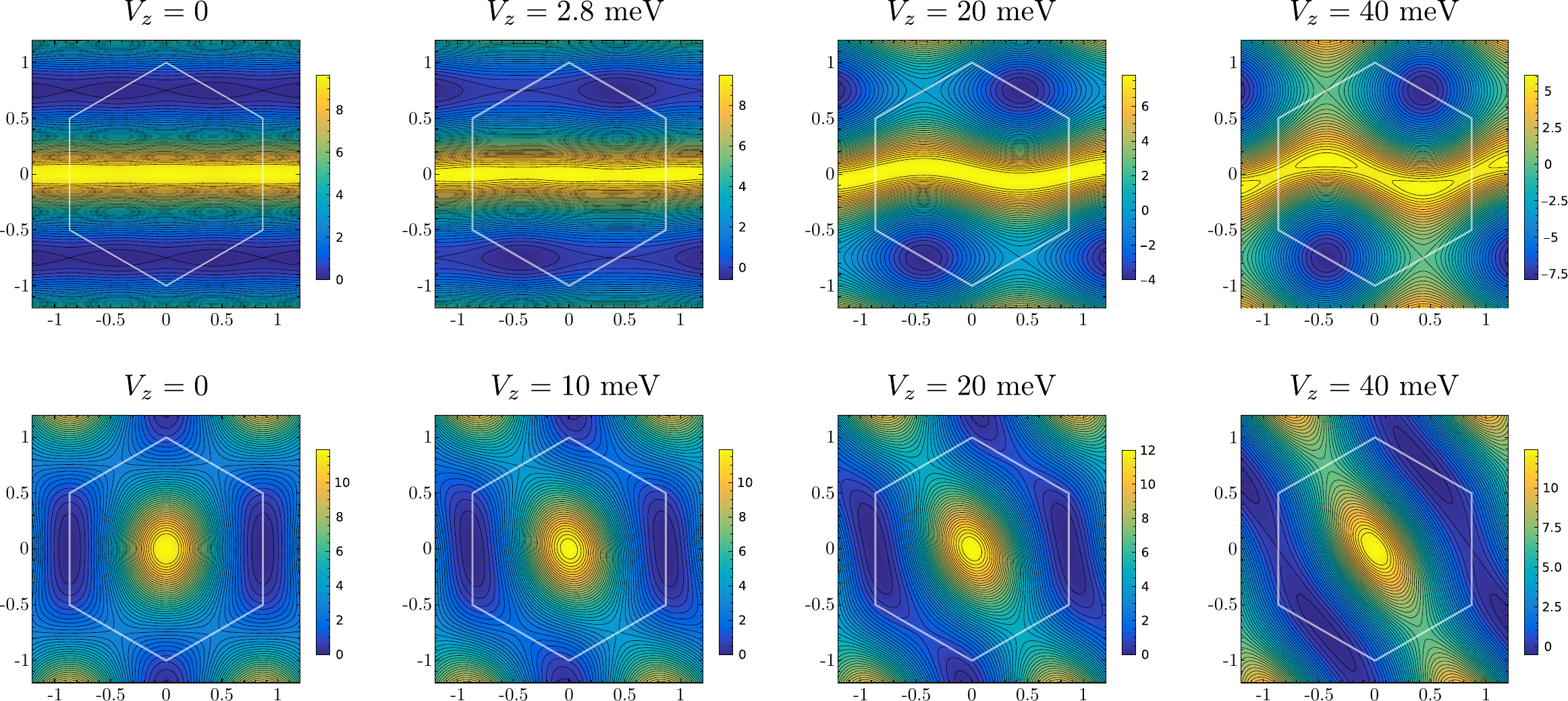}
    \caption{Energy dispersion of the lowest-energy $\tau = 1$ moir\'e band for aligned (top row) and antialigned (bottom row) stacking of $3.89^\circ$ 1T tSnSe$_2$ for several $V_z$ as indicated. For aligned (antialigned) stacking, the energy zero is set at the low-energy Van Hove singularity ($\mu_1$ point of the MBZ). As we increase $V_z$, two low-energy saddle points merge with a minimum resulting in contours that touch tangentially for $V_z \approx 2.8$ meV, giving rise to a high-order VHS. Moreover, the band evolves from quasi-1D to 2D, as a finite $V_z$ breaks $\mathcal M_z$ symmetry \cite{calugaru_moire_2025} and one eventually recovers two decoupled layers. On the other hand, for antialigned stacking the VHS at half filling for $V_z = 0$ moves towards charge neutrality as we increase $|V_z|$. Simultaneously the band evolves from 2D to quasi-1D in an intermediate $V_z$ window.}
    \label{fig:fig7}
\end{figure*}

On the other hand, for antialigned stacking, there is a VHS close to half filling ($\nu = 3$) for $V_z=0$ due to saddle points in the dispersion at the $\bm \mu_{\tau l}$ points, which are split in energy at finite $V_z$. Accordingly, the VHS moves from half filling to charge neutrality as $|V_z|$ increases, see bottom left panel of Fig.\ \ref{fig:fig6}. We also note that in a given valley, the moir\'e bands for aligned stacking are quasi one-dimensional (1D) near zero interlayer bias, but become increasingly 2D as $|V_z|$ increases. Remarkably, the opposite is true for antialigned stacking. This implies that the quasi 1D nature of the $M$-valley moir\'e bands can be tuned \textit{in situ} with an applied electric field perpendicular to the twisted layers. This is illustrated in Fig.\ \ref{fig:fig7} where we show the energy contours of the lowest-energy moir\'e band of valley $\tau = 1$ for several $V_z$.

When interactions are included, the DOS is significantly modified. The Hartree and Fock terms change the dispersion of the VHS in the $\nu-V_z$ plane from ``V shaped'' to ``Y shaped'' which results from pinning of the Fermi energy to the VHS. This phenomenon also occurs in Hubbard models \cite{irkhin_robustness_2002}, kagome metals \cite{nag_pomeranchuk_2024}, and twisted bilayer graphene \cite{cea_electronic_2019,ezzi_self-consistent_2024}. Here we find that the VHS pinning is driven by the Fock term which favors a simply-connected Fermi pocket pushing the VHS to the Fermi surface \cite{huang_spin_2023}. 
In addition, the peaks broaden such that the Lifshitz transition is washed out in the DOS. Moreover, for aligned stacking, the secondary VHS near full filling is strongly suppressed by the Hartree term which broadens the band.

While we have only considered a single twist angle in this section, we find that the density of states is qualitatively similar for twist angles $\lesssim 5^\circ$ for which the moir\'e effect is strong. For larger twist angles, the distinction between aligned and antialigned stacking is reduced and the moir\'e potentials are weaker compared to the kinetic energy scale set by the twist angle. This is evident from the Lifshitz transition $\nu \propto |V_z|$ between layer-polarized and layer-hybridized regimes, as shown in the SM \cite{Note1}.
\begin{table*}
    \label{tab:HF}
    \centering
    \begin{tabular}{c | c | c | c | c | c | c}
        \Xhline{1pt}
        & cell & $\bm Q$ & $\mathcal T$  & $\mathcal C_{3z}$ & $\left< S_z \right> / N_e$ & order parameter \\
        \hline
        M & $1 \times 1$ & $\{ \bm 0 \}$ & yes & yes & $0$ & n.a. \\
        \hline
        FM & $1 \times 1$ & $\{ \bm 0 \}$ & no & yes & $3/2$ & $S_z$ \\
        \hline
        nematic FM & $1 \times 1$ & $\{ \bm 0 \}$ & no & no & $1/2$ & $S_z$ \\
        \hline
        CDW & $2 \times 1$ & $\{\bm g_1 / 2, \bm g_1/2 + \bm g_2 \}$ & yes & no & $0$ & $\rho(\bm r) = \sum_{\bm Q} e^{i \bm Q \cdot \bm r} \rho_{\bm Q}$ \\
        \hline
        AFM & $2 \times 1$ & $\{\bm g_1 / 2, \bm g_1/2 + \bm g_2 \}$ & no & only charge density & 0 & $S_z(\bm r) = \sum_{\bm Q} e^{i \bm Q \cdot \bm r} S_{z,\bm Q}$ \\
        \Xhline{1pt}
    \end{tabular} 
    \caption{Hartree-Fock candidate ground states at half filling ($\nu = 3$) considered in this work showing the reconstructed cell in real space and the corresponding nesting vectors $\bm Q$. We also list whether the state breaks time-reversal symmetry ($\mathcal T$) or threefold rotations symmetry ($\mathcal C_{3z}$), we give the spin expectation value per particle in units of $\hbar$, and the order parameter.} 
\end{table*}

\textcolor{NavyBlue}{\textit{Correlated insulator at half-filling}} --- Finally, we consider the symmetry-broken correlated insulator at half filling ($\nu = 3$) of the low-energy moir\'e bands.  To describe these correlated states, it is convenient to introduce the single-particle density matrix $\hat{\rho}(\bm k) = \langle c_{s\tau\bm Q,\bm k}^\dag c_{s'\tau'\bm Q',\bm k} \rangle$, where $\bm k$ is restricted to the supercell Brillouin zone and $\bm Q$ labels associated reciprocal lattice vectors of the enlarged cell. In the absence of spontaneous breaking of moir\'e translational symmetry $\bm Q \in \{ \bm 0 \}$. Different Hartree–Fock symmetry-broken ansatzes correspond to different choices of $\hat{\rho}(\bm k)$, subject to the constraints $( 1/ N_\text{m} ) \sum_{\bm k}\mathrm{Tr}[\hat{\rho}(\bm k)] = \nu$ and $\hat{\rho}^2(\bm k) = \hat{\rho}(\bm k)$ which enforce charge conservation and the Pauli principle. For simplicity, and to illustrate the general principles, we only consider states that conserve valley charge conservation, but allow for broken moir\'e translational symmetry. In this case, $\hat{\rho}(\bm k)$ is diagonal in valley space. Note that this analysis excludes intervalley-coherent order. 
We then consider five HF ansatzes at half filling, that are listed in Table \ref{tab:HF}, of which two break break moir\'e translational symmetry. In particular, we consider a rectangular $2\times 1$ reconstruction which breaks $\mathcal C_{3z}$ symmetry. This is justified by the projective $\mathcal M_z$ eigenbasis for aligned stacking \cite{calugaru_moire_2025}, as discussed above, and the nesting of the Fermi surface for antialigned stacking near half filling. 

\begin{figure}[!t]
    \centering
    \includegraphics[trim={0cm 0cm 0cm 0cm},clip,width=\linewidth]{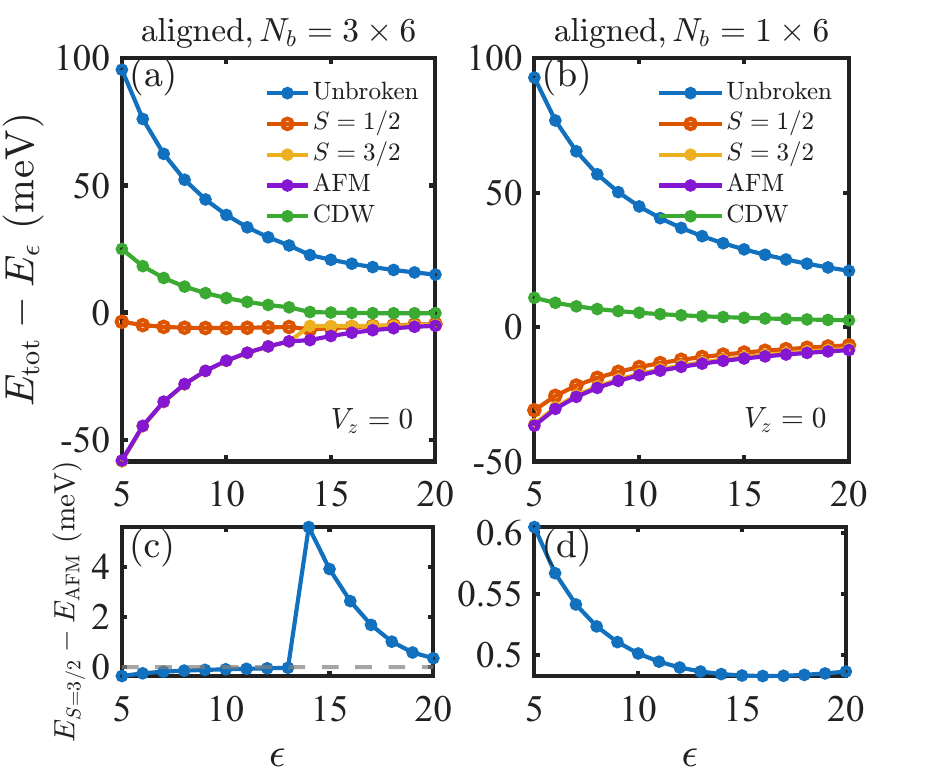}
    \caption{Self-consistent energy of different HF ansatzes (listed in Table \ref{tab:HF}) as a function of the relative dielectric constant $\epsilon$ for $3.89^\circ$ 1T tSnSe$_2$ with $V_z = 0$ for twisting near aligned (a) and antialigned (b) stacking. Here we include $N_b = 3 \times 6$ to achieve convergence. For aligned stacking, the sudden drop in energy around $\epsilon = 20$ in (a) is due to band mixing. In this case, the FM states can lower their energy by admixing the remote bands. This is further illustrated in (c) and (d) where we neglect band mixing by projecting on the low-energy moir\'e bands ($N_b = 1 \times 6$). To better visualize the energy differences between different states we subtracted a global energy shift $E_\epsilon$ which is the average energy of every HF state for each $\epsilon$.}
    \label{fig:fig8}
\end{figure}

We first discuss the three HF states that conserve moir\'e translations. These are given by: (1) the symmetry-unbroken state (M) which is metallic and preserves both time-reversal ($\mathcal T$) and threefold rotation ($\mathcal C_{3z}$) symmetry, (2) a fully spin-polarized state ($S=3/2$) that equally populates the three $M$ valleys and thus preserves $\mathcal C_{3z}$ but breaks $\mathcal T$, and (3) a partially spin-polarized ($S=1/2$) nematic state which breaks both $\mathcal T$ and $\mathcal C_{3z}$. In addition, we consider a $2\times 1$ charge-density wave (CDW) and antiferromagnet (AFM) which also break $\mathcal C_{3z}$. Figures of the charge and spin densities of each self-consistent solution are given in the SM \cite{Note1}. The energy of each self-consistent HF solution is shown in Fig.\ \ref{fig:fig8}(a) and (b) for $V_z = 0$ as a function of (inverse) interaction strength $\epsilon$ for both aligned and antialigned $3.89^\circ$ tSnSe$_2$, respectively. Out of these, the (2,1) AFM is the ground state for $V_z = 0$ except in the limit of strong interactions. Note that there are three degenerate AFMs related by $\mathcal C_{3z}$. One of these may be favored in the presence of external strain. For aligned stacking, there is a transition driven by band mixing between the low-energy manifold and the remote bands, to the isotropic FM for $\epsilon < 15$. In Fig.\ \ref{fig:fig8}(c) we show the energy in the absence of band mixing, i.e.\ by first projecting the theory on the low-energy bands. Note that this is consistent with the reduced gap of the symmetry-unbroken HF state as compared to the noninteracting case for aligned stacking, see Fig.\ \ref{fig:fig4}. On the other hand, for antialigned stacking, band renormalization driven by interactions is weaker and band mixing is small. In fact, HF corrections increase the gap for antialigned stacking at finite doping, see Fig.\ \ref{fig:fig5}.

\begin{figure}[!t]
    \centering
    \includegraphics[trim={0.3cm 0cm 0.9cm 0.8cm},clip,width=\linewidth]{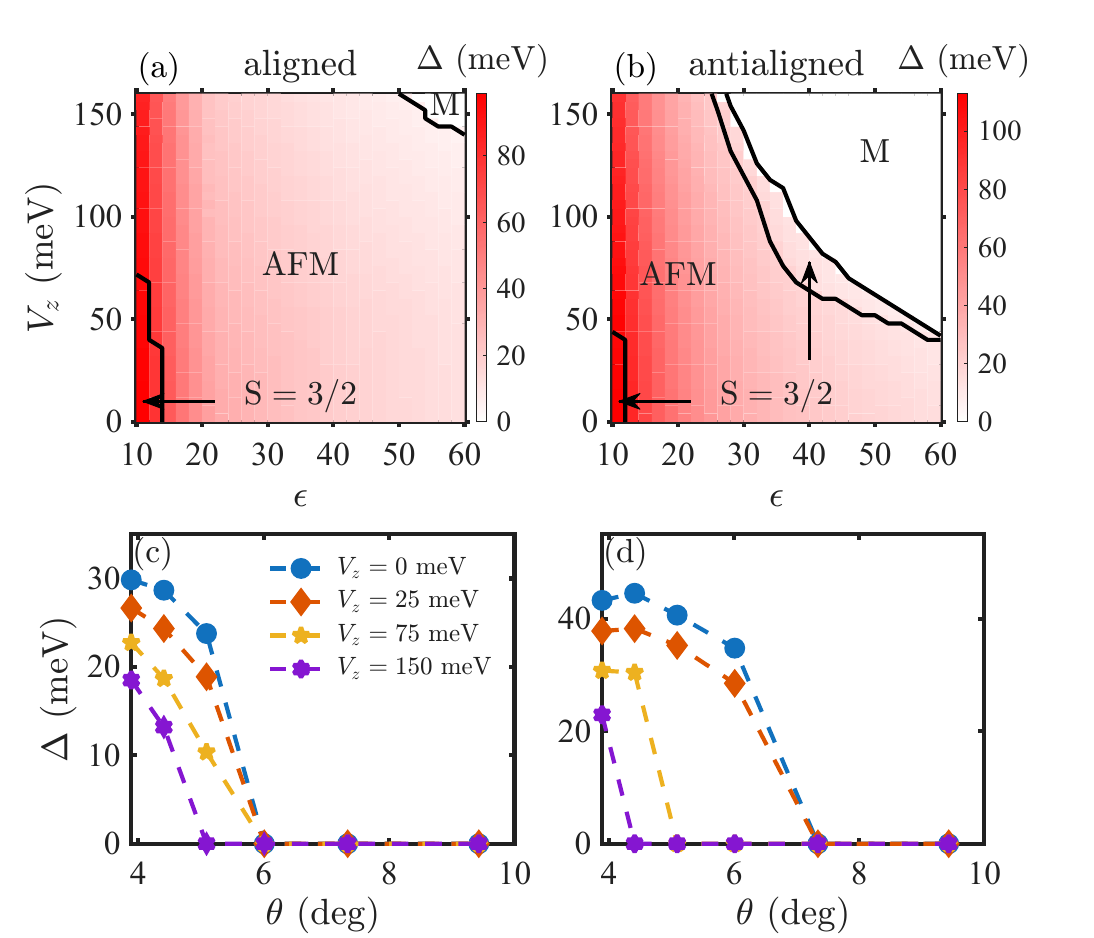}
    \caption{(top) Phase diagram and correlated gap $\Delta$ at half filling ($\nu = 3$) of the low-energy band manifold for $3.89^\circ$ 1T tSnSe$_2$ as a function of the relative dielectric constant $\epsilon$ of the encapsulating dielectric and the interlayer bias $V_z$ induced by an external electric field for (a) aligned stacking and (b) antialigned stacking. The gap generally decreases with increasing $\epsilon$ due to enhanced screening. (bottom) Correlated gap of the ground state for (c) aligned and (d) antialigned stacking as a function of twist angle $\theta$ for several $V_z$ with $\epsilon = 25$. As $\theta$ increases, the correlated gap closes and the system becomes a mixture of a gapless AFM and the symmetry-unbroken state.}
    \label{fig:fig9}
\end{figure}

We now discuss the zero-temperature phase diagram at half filling as a function of $\epsilon$ and interlayer bias $V_z$. This is shown in Fig.\ \ref{fig:fig9}(a) and (b) for aligned and antialigned $3.89^\circ$ tSnSe$_2$, respectively. Here, the color scale gives the magnitude of the correlated gap $\Delta$ and we only consider $V_z > 0$ as $V_z < 0$ is symmetry-related. For both stacking types, the ground state in the upper-right region of the phase diagram is the symmetry-unbroken metallic state, since this corresponds to weaker interactions (large $\epsilon$) and larger bandwidths (large $|V_z|$). As one moves downwards along the diagonal, the AFM becomes the ground state for aligned stacking, while for antialigned stacking there is an intermediate region given by the isotropic FM ($S=3/2$) that conserves $\mathcal C_{3z}$. While both these states break time-reversal symmetry, they can be distinguished in a transport experiment by measuring nematic order parameters such as the conductivity components $\sigma_{xx} - \sigma_{yy}$ or $\sigma_{xy} + \sigma_{yx}$ when tuned away from half filling. These combinations are only finite in the absence of threefold (and fourfold) rotation symmetry. For strong interactions and small interlayer biases (lower-left region) the system favors the FM. We further see that the correlated gap generally decreases with increasing $\epsilon$ due to enhanced screening. However, as can be seen in Fig.\ \ref{fig:fig8}, throughout most of the phase diagram, there is only a small (few meV) energy difference between the AFM and FM ground states. Hence, the actual ground state observed in experiment will be sensitive to perturbations such as external strain which may favor either one. 


Finally, we show the correlated gap as a function of twist angle for several values of $V_z$ in Fig.\ \ref{fig:fig9}(c) and (d) for $\epsilon = 25$. As expected, the AFM gap decreases when the ratio between the bandwidth, which is controlled by twist angle and $V_z$ (see Fig.\ \ref{fig:fig3}), and the interaction strength, controlled by $\epsilon$, increases. When the gap closes, the ground state is a mixture of a gapless AFM and the symmetry-unbroken state. 

\textcolor{NavyBlue}{\textit{Conclusions}} --- The possibility of experimentally exfoliable $M$-valley moir\'e systems in 1T twisted transition metal dichalcogenide homobilayers (tTMDs) fundamentally expands the landscape of correlated moir\'e physics beyond the well-explored $K/K'$ and $\Gamma$ valley platforms. The threefold valley degree of freedom can give rise to a potentially much richer phase diagram than in one ($\Gamma$) or two ($K/K'$) valley moir\'e materials, with new analogs to high-energy physics. The latter is suggested by the decomposition: $\mathrm{U(2)} \times \mathrm{U(2)} \times \mathrm{U(2)} \simeq \mathrm{U(1)} \times \mathrm{SU(2)} \times \mathrm{SU(3)}$ indicating that the $M$ valleys can be interpreted as a ``color'' charge. While we restricted ourselves here to ground states that conserve valley charge conservation, because the band dispersion breaks SU(3) explicitly, there exist a plethora of possible intervalley-coherent ground states parameterized by the Gell-Mann matrices. Fluctuations from a nascent intervalley coherent state may then serve as a pairing glue for superconductivity, similar to kagome metals \cite{zhou_chern_2022}. We have found that the color-polarized phases compete with more conventional phases seen in other materials like the FM and AFM.  This highlights the importance of external strain, which is ubiquitous in real devices, and which explicitly breaks $\mathcal C_{3z}$ symmetry, acting as an external ``color'' field.    

In particular, our demonstration that Van Hove singularities in 1T tTMDs can be pinned to the Fermi energy by electrostatic doping and external electric displacement fields opens up new pathways for engineering unconventional superconductivity and quantum criticality in van der Waals heterostructures. Furthermore, similar to other tTMDs, there is no magic angle, and the same phenomenology exists over a finite window of twist angles, making this system less sensitive to twist angle disorder. We also investigated the nature of the correlated insulator at half filling of the low-energy moir\'e bands and showed that band mixing with remote bands can be important in this moir\'e system. This is especially pronounced in aligned 1T tSnSe$_2$ for which interactions significantly reduce the gap to the remote moir\'e bands in the symmetry-unbroken HF state as we dope the low-energy moir\'e bands.  Using just electric fields, the 1T tTMDs can be tuned \textit{in situ}, between quasi-1D and 2D regimes within the same device. This creates unprecedented opportunities to study dimensional crossovers in strongly correlated systems, an important question in condensed matter physics that is difficult to study experimentally within the same device. 

The six-flavor (spin and valley) structure combined with quasi-1D channels within each $M$ valley creates opportunities for hybrid quantum phases that couple distinct many-body tendencies. We anticipate correlation-driven transitions, electrically tunable quantum criticality, and designer superconducting states with controllable gap symmetries. Unlike other moir\'e systems where such control often requires precise angle tuning, our work shows that that $M$-valley materials offer robust gate-voltage control over electronic correlations over a window of twist angles, promising more practical routes for designer quantum phases.

\let\oldaddcontentsline\addcontentsline 
\renewcommand{\addcontentsline}[3]{} 
\begin{acknowledgments}
\textcolor{NavyBlue}{\textit{Acknowledgments}} --- We thank Daniel Bennett for helpful discussions and for collaboration on another related project. Computation was done on Stampede3 at the Texas Advanced Computing Center through allocation PHY240263 from the ACCESS program supported by the U.S. National Science Foundation. L.P.\ and S.A.\ are supported by a start-up grant at Washington University in St. Louis.  C.D.B.\ and E.J.M.\ are supported by the U.S.\ Department of Energy under Grant No.\ DE-FG02-84ER45118.
\end{acknowledgments}

\textcolor{NavyBlue}{\textit{Note Added}}  --- After completion of this work, we became aware of arXiv:2508.10098v1, which also examines the role of interactions in M-valley TMDs.  In contrast to our work, this study constructs Wannier orbital models to analyze the strong-coupling limit, while our work focuses on the weak-coupling limit using self-consistent Hartree-Fock.  Together, these complimentary approaches highlight the importance of interactions across different regimes in M-valley moir\'e systems.  

\bibliography{references}
\let\addcontentsline\oldaddcontentsline 

\clearpage
\onecolumngrid
\begin{center}
\textbf{\Large Supplemental Material for: Role of electron-electron interactions in $M$-valley twisted transition metal dichalcogenides}
\end{center}
 
\setcounter{equation}{0}
\setcounter{figure}{0}
\setcounter{table}{0}
\setcounter{page}{1}
\setcounter{secnumdepth}{2}
\makeatletter
\renewcommand{\thepage}{S\arabic{page}}
\renewcommand{\thesection}{S\arabic{section}}
\renewcommand{\theequation}{S\arabic{equation}}
\renewcommand{\thefigure}{S\arabic{figure}}
\renewcommand{\thetable}{S\arabic{table}}

\tableofcontents
\vspace{1cm}

\twocolumngrid

\section{First-order Hartree energy shift}

We estimate the energy correction due to the Hartree potential using first-order perturbation theory. The Hartree potential takes the form
\begin{equation}
    H_H = \sum_{s,\tau,\mathbf{k}} \sum_{\mathbf{g} \ne 0} \rho_{\mathbf{g}} U_{\mathbf{g}}\, c^{\dagger}_{s\tau,\mathbf{k}+\mathbf{g}} c_{s\tau,\mathbf{k}},
\end{equation}
where \( \rho_{\mathbf{g}} \) is the electronic density in reciprocal space and \( U_{\mathbf{g}} \) is the Fourier component of the Coulomb interaction, given in 2D by $U_{\bm q} = e^2\tanh{(qd)}/(2\epsilon \epsilon_0 q)$ where $d$ is the gate-to-sample distance.

To compute the energy shift to a Bloch eigenstate $|\psi_{\tau n\mathbf{k}}\rangle$, we evaluate the expectation value of $H_H$ in that state
\begin{equation}
    \Delta E_{\tau n\mathbf{k}}^{(1)} = \langle \psi_{\tau n\mathbf{k}} | H_H | \psi_{\tau n\mathbf{k}} \rangle.
\end{equation}
The Bloch state in the plane-wave basis is
\begin{equation}
    |\psi_{\tau n \mathbf{k}}\rangle = \sum_{\mathbf{g}} u_{\tau n\mathbf{k}}^{(l)}(\mathbf{g})\, |\tau, \mathbf{k} + \mathbf{g} \rangle.
\end{equation}
Using this, we obtain
\begin{equation}
    \Delta E_{\tau n \mathbf{k}}^{(1)} = \sum_{\mathbf{g} \ne 0} \rho_{\mathbf{g}} U_{\mathbf{g}} \sum_{\mathbf{g}'} \left[ u_{\tau n \mathbf{k}}^{(l)}(\mathbf{g}') \right]^* u_{\tau n \mathbf{k}}^{(l)}(\mathbf{g}' + \mathbf{g}).
\end{equation}

We define the form factor of the state as
\begin{equation}
    \Lambda_{\tau n \mathbf{k}}(\mathbf{g}) = \sum_{l,\mathbf{g}'} \left[ u_{\tau n \mathbf{k}}^{(l)}(\mathbf{g}') \right]^* u_{\tau n \mathbf{k}}^{(l)}(\mathbf{g}' + \mathbf{g}),
\end{equation}
so that the energy shift becomes
\begin{equation}
    \Delta E_{\tau n \mathbf{k}}^{(1)} = \sum_{\mathbf{g} \neq 0} \rho_{\mathbf{g}} U_{\mathbf{g}}\, \Lambda_{\tau n \mathbf{k}}(\mathbf{g}).
\end{equation}

This expression provides a direct estimate of the Hartree-induced level shift in the band structure, based on the plane-wave coefficients of the Bloch states and the density distribution.

\section{Angle-dependent fitting parameters of the continuum model}

This section we provide the numerical values of the angle-dependent parameters used in the continuum model calculations presented in the main text. Table~\ref{tab:fitting_params} lists the effective masses $m_x$ and $m_y$, the intralayer moir\'e potential amplitudes $v_{11}$, $v_{12}$, and $v_{13}$ together with their phases $\psi_{11}$, $\psi_{12}$, and $\psi_{13}$, as well as the interlayer tunneling parameters $w_1$ and $w_2$ with phases $\phi_1$ and $\phi_2$. The effective masses $m_x$ and $m_y$ are given in units of the bare electron mass $m_e$, and the moir\'e potential amplitudes are in the unit of meV. The definitions of the intralayer moir\'e potential terms follow Eq.~\eqref{eq:V1l} of the main text, while the interlayer tunneling terms are defined in Eq.~\eqref{eq:T1}.  

\setlength{\tabcolsep}{3pt}
\begin{table*}[htbp]
    \centering
    \renewcommand{\arraystretch}{1.2}
    \begin{tabular}{c c c c c c c c c c c c c c}
        \toprule
        Stacking & $\theta$ & $m_x$ & $m_y$ & $v_{11}$ & $\psi_{11}$ & $v_{12}$ & $\psi_{12}$ & $v_{13}$ & $\psi_{13}$ & $w_1$ & $\phi_1$ & $w_2$ & $\phi_2$ \\
        \hline
        AA & $3.89^\circ$ & $0.21$ & $0.73$ & $29.4$ & $-22.05^\circ$ & $7.27$ & $-38.24^\circ$ & $7.27$ & $-38.24^\circ$ & $110.87$ & $36.78^\circ$ & $-7.99$ & $0^\circ$ \\
        AA & $4.41^\circ$ & $0.21$ & $0.73$ & $30.29$ & $-28.13^\circ$ & $5.58$ & $-31.81^\circ$  & $5.58$ & $-31.81^\circ$ & $108.67$ & $37.97^\circ$ & $-11.06$ & $0^\circ$ \\
        AA & $5.09^\circ$ & $0.21$ & $0.73$ & $29.03$ & $-31^\circ$ & $5.92$ & $-60.45^\circ$ & $5.92$ & $-60.45^\circ$ & $104.62$ & 38.81$^\circ$ & $-9.62$ & $0^\circ$ \\
        AA & $6.01^\circ$ & $0.21$ & $0.73$ & $33.28$ & $-30.92^\circ$ & $4.96$ & $-23.05^\circ$ & $4.96$ & $-23.05^\circ$ & $101.445$ & $39.91^\circ$ & $-9.61$ & $0^\circ$ \\
        AA & $7.34^\circ$ & $0.21$ & $0.73$ & $43.79$ & $-45.36^\circ$ & $3.73$ & $-8.48^\circ$ & $3.73$ & $-8.48^\circ$ & $97.44$ & $40.85^\circ$ & $-7.92$ & $0^\circ$ \\
        AA & $9.43^\circ$ & $0.21$ & $0.73$ & $56.215$ & $-51.99^\circ$ & $13.29$ & $-148.47^\circ$ & $13.29$ & $-148.47^\circ$ & $93.43$ & $45.13^\circ$ & $-10.74$ & $0^\circ$ \\
        AB & $3.89^\circ$ & $0.24$ & $0.72$ & $-20.94$ & $0^\circ$ & $35.27$ & $173.31^\circ$ & $35.27$ & $-6.69^\circ$ & $-62.10$ & $0^\circ$ & $31.80$ & $0^\circ$ \\
        AB & $4.41^\circ$ & $0.25$ & $0.77$ & $0$ & $0^\circ$ & $33.90$ & $-22.46^\circ$ & $33.90$ & $157.54^\circ$ & $-90.78$ & $0^\circ$ & $34.38$ & $25.54^\circ$ \\
        AB & $5.09^\circ$ & $0.25$ & $0.69$ & $0$ & $0^\circ$ & $26.98$ & $156.61^\circ$ & $26.98$ & $-23.39^\circ$ & $-88.65$ & $0^\circ$ & $31.43$ & $-21.72^\circ$ \\
        AB & $6.01^\circ$ & $0.26$ & $0.79$ & $0$ & $0^\circ$ & $20.39$ & $-25.72^\circ$ & $20.39$ & $154.28^\circ$ & $-78.22$ & $0^\circ$ & $27.41$ & $20.346^\circ$ \\
        AB & $7.34^\circ$ & $0.21$ & $0.67$ & $-30.92$ & $0^\circ$ & $24.03$ & $154.80^\circ$ & $24.03$ & $25.20^\circ$ & $-79.41$ & $0^\circ$ & $26.93$ & $0^\circ$ \\
        AB & $9.43^\circ$ & $0.21$ & $0.71$ & $0$ & $0^\circ$ & $10.11$ & $36.836^\circ$ & $11.35$ & $135.46^\circ$ & $-99.12$ & $0^\circ$ & $17.85$ & $0^\circ$ \\
        \hline
    \end{tabular}
    \caption{Angle-dependent fitting parameters of the moir\'e continuum model for twisted bilayer SnSe\(_2\). The effective masses $m_x$, $m_y$ are given in units of the bare electron mass $m_e$. The intralayer moir\'e potential amplitudes $v_{11}$, $v_{12}$, $v_{13}$ and phases $\psi_{11}$, $\psi_{12}$, $\psi_{13}$ are defined as in Eq.~\eqref{eq:V1l} of the main text. The interlayer tunneling parameters $w_1$, $w_2$ and their respective phases $\phi_1$, $\phi_2$ enter the tunneling term Eq.~\eqref{eq:T1}. These parameters are obtained by symmetry-constrained fits to DFT calculations as described in Ref.~\cite{calugaru_moire_2025}.}
    \label{tab:fitting_params}
\end{table*}

\section{Hartree-Fock Theory}

In this section we set up the Coulomb interaction and derive the Hartree-Fock framework used in our calculations. Our theoretical framework is based primarily on the many-body perturbation approach for moir\'e systems from Ref.\ \cite{peng_many-body_2025,peng_magnetism_2025}. We start from the total Hamiltonian $\mathcal{H} = \mathcal{H}_{0} + \mathcal{H}_{\mathrm{I}}$. The non-interacting Hamiltonian is given by
\begin{equation}
\mathcal{H}_{0}
=\sum_{\tilde{\mathbf{k}},\tilde{\mathbf{k}}^{\prime}}\sum_{s,\tau}  \mathcal{H}^{\tau}(\tilde{\mathbf{k}},\tilde{\mathbf{k}}^{\prime}) \hat{c}^{\dagger}_{s,\tau,\tilde{\mathbf{k}}} \hat{c}_{s,\tau,\tilde{\mathbf{k}}^{\prime}},
\end{equation}
with electron creation and annihilation operators $\hat{c}^{\dagger}_{s,\tau,\tilde{\mathbf{k}}}$ and $\hat{c}_{s,\tau,\tilde{\mathbf{k}}}$ for momentum $\tilde{\mathbf{k}}$, spin $s$, and valley $\tau$. The interacting Hamiltonian takes the form
\begin{equation}
\begin{aligned}
\mathcal{H}_{\mathrm{I}}&=
\frac{1}{2}
\sum_{\tilde{\mathbf{k}},\tilde{\mathbf{k}}^{\prime},\tilde{\mathbf{q}}}
\sum_{s,s^{\prime}}
\sum_{\tau,\tau^{\prime}}
V_{\tilde{\mathbf{q}}}
\hat{c}_{s,\tau,\tilde{\mathbf{k}}+\tilde{\mathbf{q}}}^{\dagger}
\hat{c}_{s^{\prime},\tau^{\prime},\tilde{\mathbf{k}}^{\prime}-\tilde{\mathbf{q}}}^{\dagger}
\hat{c}_{s^{\prime},\tau^{\prime},\tilde{\mathbf{k}}^{\prime}}
\hat{c}_{s,\tau,\tilde{\mathbf{k}}},\\
\end{aligned}
\end{equation}
where momenta $\tilde{\mathbf{k}},\tilde{\mathbf{k}}^{\prime},\tilde{\mathbf{q}}$ lie within the Brillouin zone.

To handle translation symmetry-breaking orders, we introduce enlarged unit cells, or equivalently fold the moir\'e Brillouin zone \cite{xie_phase_2023}. Each order corresponds to a specific reciprocal lattice vector pair $\mathbf{Q}_{1,2}$. The folding number is defined as
\begin{equation} N_F=\frac{|\mathbf{g}_1\times \mathbf{g}_2|}{|\mathbf{Q}_1\times \mathbf{Q}_2|}, \end{equation} 
which represents the number of times the Brillouin zone is folded, where $\mathbf{g}_i$ are reciprocal lattice vectors of the mBZ. Any momentum $\tilde{\mathbf{k}}$ in the BZ can then be decomposite as
\begin{equation}
\tilde{\mathbf{k}}=\mathbf{k}+\mathbf{Q}+\mathbf{g},
\end{equation}
where $\mathbf{Q}=l_1\mathbf{Q}_1+l_2\mathbf{Q}_2$ with $l_1$ and $l_2$ are integer number stand for all the $N_F$ reciprocal vectors of the folded moir\'e Brillouin zones in the unfolded moir\'e Brillouin zones. Specifically, for FM and symmetry unbroken states, we have $\mathbf{Q}_1=\mathbf{g}_1$, $\mathbf{Q}_2=\mathbf{g}_2$ and $N_F=1$. For the CDW and AFM states, $\mathbf{Q}_1=\mathbf{g}_1/2$, $\mathbf{Q}_2=\mathbf{g}_2$ and $N_F=2$.  Using this decomposition, the non-interacting part reads
\begin{equation}
\mathcal{H}_{0}=\sum_{s,\tau,n,\mathbf{k},\mathbf{Q}} E^{s}_{\tau,n}(\mathbf{k},\mathbf{Q})
\hat{c}^{\dagger}_{s,\tau,n,\mathbf{k},\mathbf{Q}} \hat{c}_{s,\tau,n,\mathbf{k},\mathbf{Q}} .
\end{equation}
Here, $\hat{c}^{\dagger}_{s,\tau,n,\mathbf{k},\mathbf{Q}}$ ($\hat{c}_{s,\tau,n,\mathbf{k},\mathbf{Q}}$) creates (annihilates) an electron with momentum $\mathbf{k}+\mathbf{Q}$, spin $s$, valley $\tau$, and band index $n$. The band dispersion is denoted by $E^{s}_{\tau,n}(\mathbf{k},\mathbf{Q})$.  

The interaction part of the Hamiltonian is given by
\begin{widetext}
\begin{equation}
\begin{aligned}
\mathcal{H}_{\mathrm{I}}=\frac{1}{2}
\sum_{\{n_i\}}
\sum_{\mathbf{k},\mathbf{k}^{\prime},\mathbf{q}}
\sum_{\mathbf{Q},\mathbf{Q}^{\prime}}
\sum_{s,s'}
\sum_{\tau,\tau'}
V^{\tau\tau^{\prime}}_{\mathbf{k},\mathbf{k}^{\prime},\mathbf{Q},\mathbf{Q}^{\prime},\mathbf{q},\{n_i\}}
\hat{c}_{s,\tau,n_1,\mathbf{k}+\mathbf{q},\mathbf{Q}+\mathbf{Q}^{\prime\prime}}^{\dagger}
\hat{c}_{s',\tau',n_2,\mathbf{k}^{\prime}-\mathbf{q},\mathbf{Q}^{\prime}-\mathbf{Q}^{\prime\prime}}^{\dagger}
\hat{c}_{s',\tau',n_3,\mathbf{k}^{\prime},\mathbf{Q}^{\prime}}
\hat{c}_{s,\tau,n_4,\mathbf{k},\mathbf{Q}},
\end{aligned}
\end{equation}
with the matrix element
\begin{equation}
\begin{aligned}
V^{\tau\tau^{\prime}}_{\mathbf{k},\mathbf{k}^{\prime},\mathbf{Q},\mathbf{Q}^{\prime},\mathbf{q},\{n_i\}}=
\sum_{\mathbf{g},\mathbf{Q}^{\prime\prime}}
V_{\mathbf{q}+\mathbf{Q}^{\prime\prime}+\mathbf{g}}
\left[\Lambda^{*}_{\mathbf{k}+\mathbf{Q},\mathbf{q}+\mathbf{Q}^{\prime\prime}+\mathbf{g}}\right]^{\tau}_{n_4n_1}
\left[\Lambda^{*}_{\mathbf{k}^{\prime}+\mathbf{Q}^{\prime},-\mathbf{q}-\mathbf{Q}^{\prime\prime}-\mathbf{g}}\right]^{\tau^\prime}_{n_3n_2},
\end{aligned}
\end{equation}
where the form factor is defined as
\begin{equation}
\left[\Lambda_{\mathbf{k}+\mathbf{Q},\mathbf{q}+\mathbf{Q}^{\prime\prime}+\mathbf{g}}\right]^{\tau}_{m n}=\sum_{\mathbf{g}^{\prime}}
u^*_{m} \left(\mathbf{k}+\mathbf{Q}+\mathbf{g}^{\prime} ; \tau\right) 
u_{n} \left(\mathbf{k}+\mathbf{Q}+\mathbf{g}^{\prime}    +\mathbf{q}+\mathbf{Q}^{\prime\prime}+\mathbf{g} ; \tau\right).
\end{equation}
\end{widetext}
In the band basis, the interaction Hamiltonian resembles the standard Coulomb interaction but with modified interaction strengths that depend on the form factors and band indices.

We then define the single-particle imaginary-time Green's function in the band basis
\begin{equation}
    \left[G(\mathbf{k},\tau)\right]_{\eta\eta^{\prime}}=-\langle \mathcal{T}_{\tau} \hat{c}_{\mathbf{k},\eta}(\tau)\hat{c}^{\dagger}_{\mathbf{k},\eta^{\prime}}(0) \rangle,
\end{equation}
where $\eta=(n,s,\tau,\bm Q)$. Here, $\hat{c}_{\mathbf{k},\eta}(\tau)$ and $\hat{c}^{\dagger}_{\mathbf{k},\eta^{\prime}}(0)$ are the annihilation and creation operators in the Heisenberg picture, $\mathcal{T}_{\tau}$ denotes the time-ordering operator, and $\langle \cdots \rangle$ represents the thermal average over the interacting system. By performing a Fourier transform with respect to imaginary time $\tau$, we obtain the Green's function in frequency space
\begin{equation}
\left[G(\mathbf{k},i\omega_n)\right]_{\eta\eta^{\prime}}=\int_{0}^{\beta}e^{i\omega_n\tau}\left[G(\mathbf{k},\tau)\right]_{\eta\eta^{\prime}}d\tau,
\end{equation}
where $\beta = 1/(k_{\mathrm{B}} T)$ is the inverse temperature, $k_{\mathrm{B}}$ is Boltzmann's constant, and $i\omega_n = (2n + 1)\pi/\beta$ are the fermionic Matsubara frequencies.

For non-interacting electrons the propagator is diagonal
\begin{equation}
\left[G_{0}(\mathbf{k},i\omega_n)\right]_{\eta\eta^{\prime}} = \frac{\delta_{\eta,\eta^{\prime}}}{i\omega_n - E^{s}_{\tau,n}(\mathbf{k},\mathbf{Q}) +\mu}, 
\end{equation} 
where $E^{s}_{\tau,n}(\mathbf{k},\mathbf{Q})$ are the eigenvalues of the single-particle Hamiltonian, and $\mu$ is the global chemical potential. The interacting Green's function, $\hat{G}(\mathbf{k},i\omega_n)$, is determined by the Dyson equation
\begin{equation}\label{eq:Dyson}
\hat{G}^{-1}(\mathbf{k},i\omega_n) = \hat{G}^{-1}_0(\mathbf{k},i\omega_n) - \hat{\Sigma}(\mathbf{k},i\omega_n),
\end{equation}
where $\hat{\Sigma}(\mathbf{k},i\omega_n)$ represents the electron self-energy, and $\hat{G}_0(\mathbf{k},i\omega_n)$ denotes the non-interacting Green's function. Using Feynman diagram techniques, the Hartree self-energy could be written down as
\begin{widetext}
\begin{equation}
\begin{aligned}
    \left[\Sigma_{\mathrm{H}}(\mathbf{k}^{\prime},i\omega_n)\right]_{\mathbf{Q}^{\prime},n_3,s^{\prime},\tau^{\prime}}^{\mathbf{Q}^{\prime}-\mathbf{Q}^{\prime\prime},n_2,s^{\prime},\tau^{\prime}}& = \sum_{\mathbf{g}^{\prime\prime}} V_{\mathbf{Q}^{\prime\prime}+\mathbf{g}^{\prime\prime}} \left[\Lambda_{\mathbf{k}^{\prime}+\mathbf{Q}^{\prime},\mathbf{Q}^{\prime\prime}+\mathbf{g}^{\prime\prime}}\right]^{\tau^\prime}_{n_2n_3} \\
    & \times \sum_{\mathbf{k}} \frac{1}{\beta}\sum_{m} \sum_{\mathbf{Q}} \sum_{n_1,n_4,s,\tau} \left[\Lambda^{*}_{\mathbf{k}+\mathbf{Q},\mathbf{Q}^{\prime\prime}+\mathbf{g}^{\prime\prime}}\right]^{\tau}_{n_4n_1} \left[G(\mathbf{k},i\omega_m)\right]^{\mathbf{Q},n_4,s,\tau}_{\mathbf{Q}+\mathbf{Q}^{\prime\prime},n_1,s,\tau},\\
\end{aligned}
\end{equation}
Similarly, for Fock self-energy
\begin{equation}
\begin{aligned}
&\left[\Sigma_{\mathrm{F}}(\mathbf{k},i\omega_n)\right]^{\mathbf{Q}^{\prime},n_2,s^{\prime},\tau^{\prime}}_{\mathbf{Q},n_4,s,\tau}\\
&=-\frac{1}{\beta}\sum_{m}
\sum_{n_1,n_3}
\sum_{\mathbf{q},\mathbf{g}^{\prime\prime},\mathbf{Q}^{\prime\prime}}
V_{\mathbf{q}+\mathbf{Q}^{\prime\prime}+\mathbf{g}^{\prime\prime}}
\left[\Lambda^{*}_{\mathbf{k}+\mathbf{Q},\mathbf{q}+\mathbf{Q}^{\prime\prime}+\mathbf{g}^{\prime\prime}}\right]^{\tau}_{n_4n_1}
\left[G(\mathbf{k}+\mathbf{q}+\mathbf{Q}^{\prime\prime},i\omega_n-i\omega_m)\right]^{\mathbf{Q}^{\prime},n_3,s^{\prime},\tau^{\prime}}_{\mathbf{Q},n_1,s,\tau}
\left[\Lambda_{\mathbf{k}+\mathbf{Q}^{\prime},\mathbf{q}+\mathbf{Q}^{\prime\prime}+\mathbf{g}^{\prime\prime}}\right]^{\tau^\prime}_{n_2n_3},\\
\end{aligned}
\end{equation}
where we have used the relationship
$\hat{\Lambda}_{\mathbf{k}+\mathbf{q},-\mathbf{q}-\mathbf{g}}=\hat{\Lambda}^{\dagger}_{\mathbf{k},\mathbf{q}+\mathbf{g}}$.

Carrying out the Matsubara summation connects the self-energy to the equal-time density matrix
\begin{equation}
\hat{\rho}(\mathbf{k})=\hat{G}(\mathbf{k},\tau=0^-)
=\frac{1}{\beta}\sum_{n}e^{-i\omega_n0^+}\hat{G}(\mathbf{k},i\omega_n).
\end{equation}
In terms of $\hat{\rho}(\mathbf{k})$, the Hartree and Fock contributions reduce to
\begin{equation}\label{eq:hartree}
\begin{aligned}
\left[\Sigma_{\mathrm{H}}(\mathbf{k})\right]_{\mathbf{Q},n_3,s,\tau}^{\mathbf{Q}-\mathbf{Q}^{\prime\prime},n_2,s,\tau}
&= 
\sum_{\mathbf{g}^{\prime\prime}}
V_{\mathbf{Q}^{\prime\prime}+\mathbf{g}^{\prime\prime}}
\left[\Lambda_{\mathbf{k}+\mathbf{Q},\mathbf{Q}^{\prime\prime}+\mathbf{g}^{\prime\prime}}\right]^{\tau}_{n_2n_3}
\sum_{\mathbf{k}^{\prime},\mathbf{Q}^{\prime}}
\sum_{n_1,n_4,\tau^{\prime}}
\left[\Lambda^{*}_{\mathbf{k}^{\prime}+\mathbf{Q}^{\prime},\mathbf{Q}^{\prime\prime}+\mathbf{g}^{\prime\prime}}\right]^{\tau^{\prime}}_{n_4n_1}
\left[\rho(\mathbf{k}^{\prime})\right]^{\mathbf{Q}^{\prime},n_4,s^{\prime},\tau^{\prime}}_{\mathbf{Q}^{\prime}+\mathbf{Q}^{\prime\prime},n_1,s^{\prime},\tau^{\prime}},\\
\end{aligned}
\end{equation}
and
\begin{equation}\label{eq:fock}
\begin{aligned}
\left[\Sigma_{\mathrm{F}}(\mathbf{k})\right]^{\mathbf{Q}^{\prime},n_2,s^{\prime},\tau^{\prime}}_{\mathbf{Q},n_4,s,\tau}
&=\sum_{n_1,n_3}
\sum_{\mathbf{q},\mathbf{g}^{\prime\prime},\mathbf{Q}^{\prime\prime}}
V_{\mathbf{q}+\mathbf{Q}^{\prime\prime}+\mathbf{g}^{\prime\prime}}
\left[\Lambda^{*}_{\mathbf{k}+\mathbf{Q},\mathbf{q}+\mathbf{Q}^{\prime\prime}+\mathbf{g}^{\prime\prime}}\right]^{\tau}_{n_4n_1}
\left[\rho(\mathbf{k}+\mathbf{q}+\mathbf{Q}^{\prime\prime})\right]^{\mathbf{Q}^{\prime},n_3,s^{\prime},\tau^{\prime}}_{\mathbf{Q},n_1,s,\tau}
\left[\Lambda_{\mathbf{k}+\mathbf{Q}^{\prime},\mathbf{q}+\mathbf{Q}^{\prime\prime}+\mathbf{g}^{\prime\prime}}\right]^{\tau^\prime}_{n_2n_3}.\\
\end{aligned}
\end{equation}
\end{widetext}

In practical calculations, we impose a cutoff in the band basis denoted by $N$. The Green’s function then has dimension $(6NN_F)\times(6NN_F)$, accounting for spin and valley. In this study we use $N=3$ and verify that increasing this cutoff does not qualitatively change our results.

The Hartree-Fock equations are solved iteratively: starting from a trial mean self-energy $\hat{\Sigma}_0(\mathbf{k})$, we update the Green’s function via Dyson’s equation, compute the density matrix $\hat{\rho}(\mathbf{k})$, and reconstruct a new self-energy $\hat{\Sigma}_{\mathrm{H}}(\mathbf{k})$ and $\hat{\Sigma}_{\mathrm{F}}(\mathbf{k})$ until a fixed point is reached. As an initial guess, we populate the first moir\'e band with density matrices corresponding to different candidate phases. For the $S=1/2$ phase, we take
\begin{equation}
\hat{\Sigma}_0(\mathbf{k}) =
\alpha\begin{pmatrix}
\hat{\mathbb{I}}_2 & \hat{0} & \hat{0} \\
\hat{0} & \hat{\sigma}_z & \hat{0} \\
\hat{0} & \hat{0} & \hat{0} \\
\end{pmatrix},
\end{equation}
while for the $S=3/2$ phase we choose $\hat{\Sigma}_0(\mathbf{k})=\alpha\hat{\mathbb{I}}_3\otimes\hat{\sigma}_z$ expressed in the basis (valley $\otimes$ spin), where $\hat{\mathbb{I}}_n$ is the $n$-dimensional identity matrix and $\alpha$ is a energy constant. For the CDW and AFM state, we double the moire unit cell and the dimension of Green's function. For the CDW state, we choose the initial ansatz $\hat{\Sigma}_0(\mathbf{k})=\alpha\hat{\mathbb{I}}_3\otimes\hat{\mathbb{I}}_2\otimes\hat{\sigma}_x$ on the basis (valley $\otimes$ spin $\otimes$  $\bm Q$) while for AFM state we choose
$\hat{\Sigma}_0(\mathbf{k})=\alpha\hat{\mathbb{I}}_3\otimes\hat{\sigma}_z\otimes\hat{\sigma}_x$

Finally, the ground-state energy is computed using the Galitskii–Migdal expression \cite{galitskii_application_1958},
\begin{equation}
E_{\mathrm{tot}}
=\frac{1}{\beta}\sum_{\mathbf{k},n}e^{-i\omega_n0^+}\,
\mathrm{Tr}\Big[\big(\hat{H}_0(\mathbf{k})
+\tfrac{1}{2}\hat{\Sigma}(\mathbf{k},i\omega_n)\big)\,
\hat{G}(\mathbf{k},i\omega_n)\Big].
\end{equation}
When the self-energy is frequency-independent, this reduces to the simpler form
\begin{equation}
E_{\mathrm{tot}}=
\sum_{\mathbf{k}}
\mathrm{Tr}\Big[\big(\hat{H}_0(\mathbf{k})
+\tfrac{1}{2}\hat{\Sigma}(\mathbf{k})\big)\hat{\rho}(\mathbf{k})\Big].
\end{equation}

\section{Density of states for additional twist angles}

\begin{figure*}[!t]
    \centering
    \includegraphics[trim={0cm 0 0cm 0},clip,width=1.9\columnwidth]{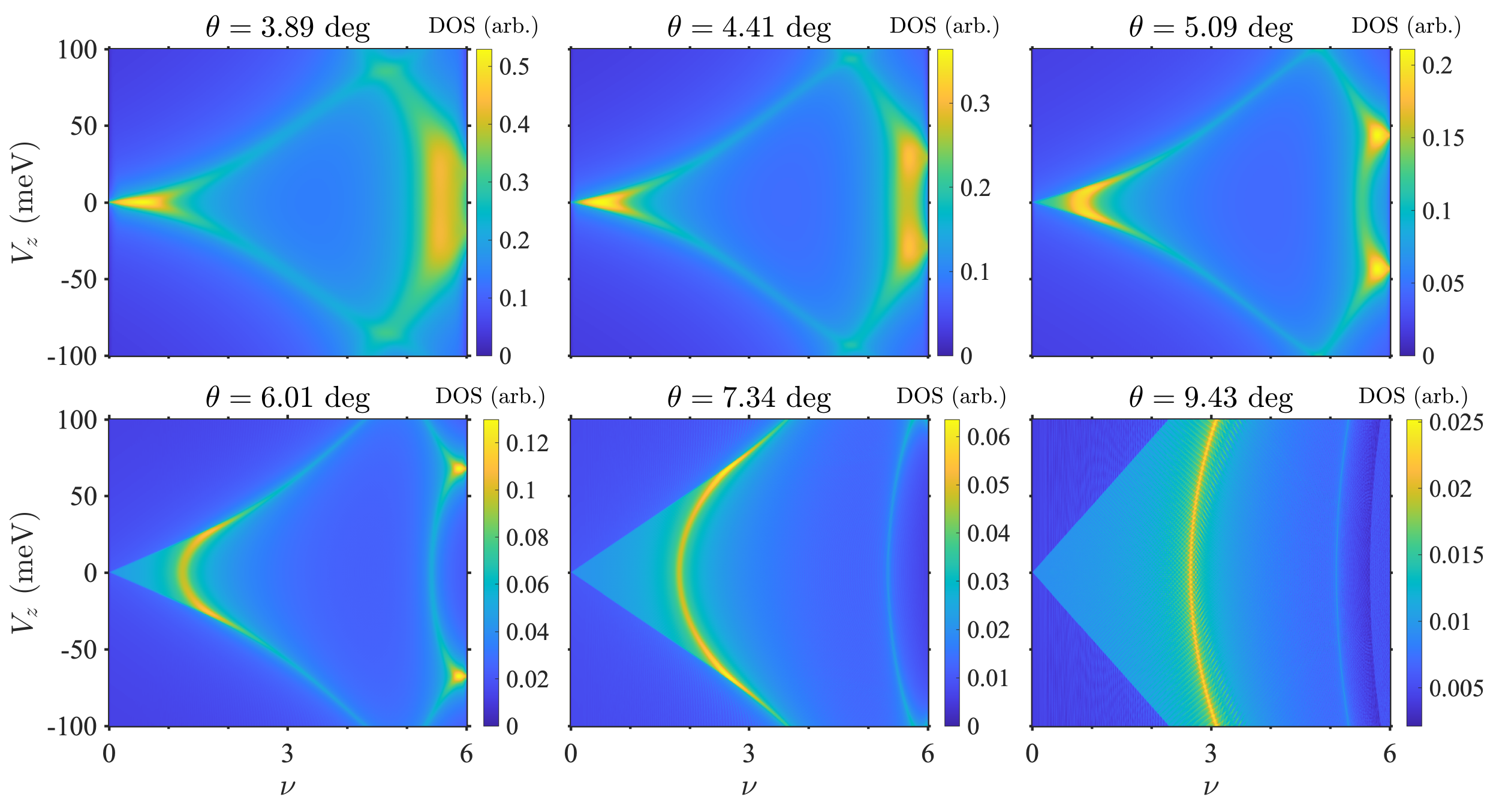}
    \caption{Density of states of the non-interacting model for aligned SnSe$_2$ at different twist angles. For twist angles $\theta \lesssim 5^\circ$, where the moir\'e effect is strong, the density of states shows qualitatively similar features. At larger twist angles, the difference between aligned and antialigned stacking is reduced, since the moir\'e potential becomes weak compared to the kinetic energy scale set by the twist. This behavior is consistent with the Lifshitz transition $\nu \propto |V_z|$ between layer-polarized and layer-hybridized regimes.}
    \label{fig:figS1}
\end{figure*}

\begin{figure*}[!t]
    \centering
    \includegraphics[trim={0cm 0 0cm 0},clip,width=1.9\columnwidth]{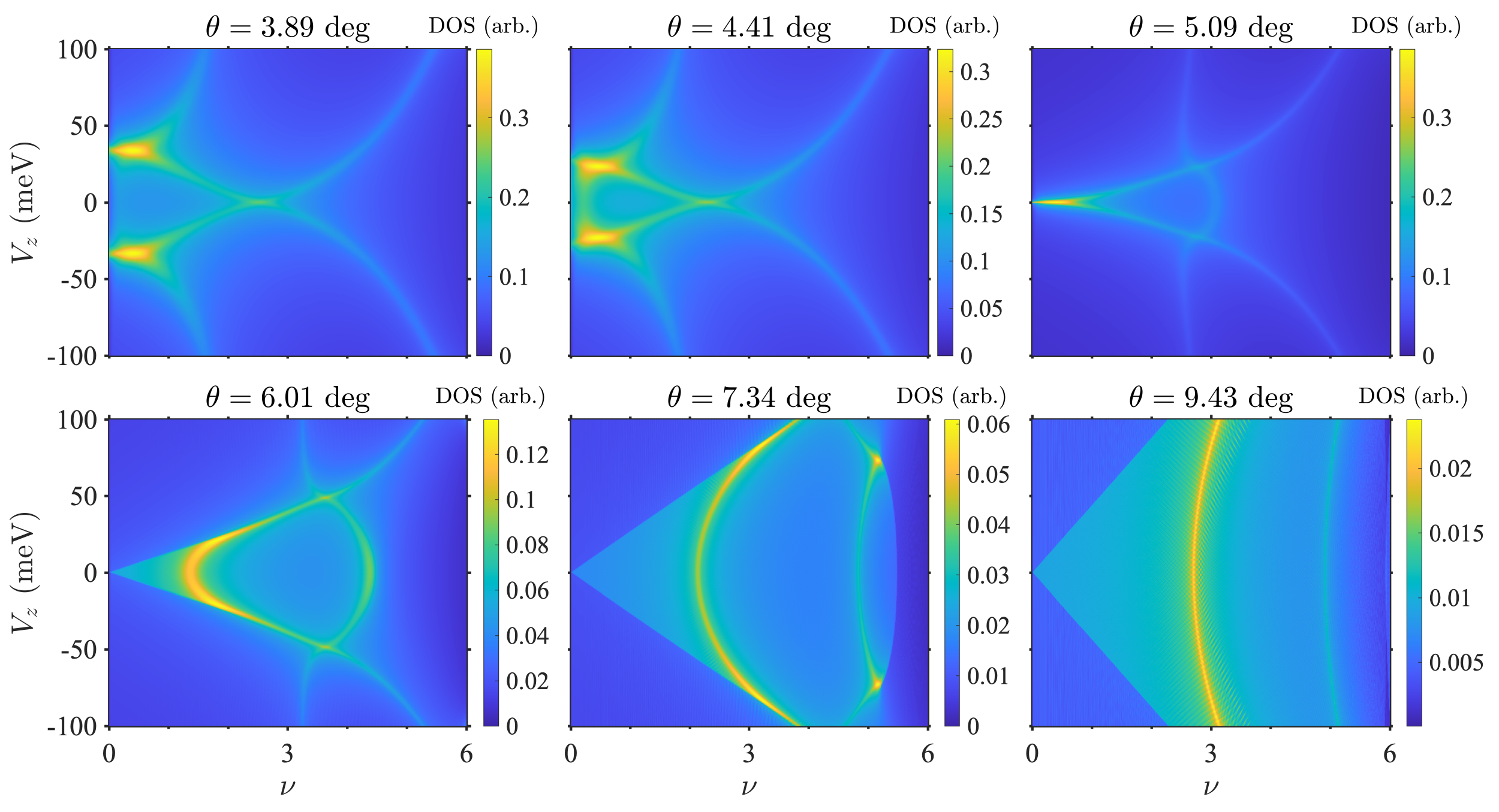}
    \caption{Density of states of the non-interacting model for antialigned SnSe$_2$ at different twist angles. For twist angles $\theta \lesssim 5^\circ$, where the moir\'e effect is strong, the density of states shows qualitatively similar features. At larger twist angles, the difference between aligned and antialigned stacking is reduced, since the moir\'e potential becomes weak compared to the kinetic energy scale set by the twist. This behavior is consistent with the Lifshitz transition $\nu \propto |V_z|$ between layer-polarized and layer-hybridized regimes.}
   \label{fig:figS2}
\end{figure*}

In the main text, we focused on a representative twist angle. Here we present the density of states for a range of twist angles and both stacking configurations (aligned and antialigned). The results are shown in Fig.~\ref{fig:figS1} for aligned and Fig.~\ref{fig:figS2} for antialigned SnSe$_2$.  

For small twist angles $\lesssim 5^\circ$, the moir\'e potential strongly modifies the band structure, leading to characteristic features in the DOS such as sharp peaks associated with flat or narrow bands. The overall shapes of the DOS are qualitatively similar across different small angles, reflecting the dominance of the moir\'e potential over the kinetic energy scale set by the twist.  

At larger twist angles, the moir\'e potential becomes weak compared to the kinetic energy. In this regime, the difference between aligned and antialigned stacking is greatly reduced, and the DOS gradually approaches that of an anisotropic two-dimensional electron gas. This crossover is consistent with the Lifshitz transition discussed in the main text, where the Fermi surface evolves between layer-polarized and layer-hybridized states as a function of $V_z$.  

Thus, the angle dependence of the DOS confirms that strong moir\'e features appear only at relatively small twist angles, while at larger angles the low-energy properties are essentially determined by the parent band anisotropy with only weak stacking dependence.


\begin{figure}[!t]
    \centering
    \includegraphics[trim={0cm 0cm 0cm 0cm},clip,width=1\columnwidth]{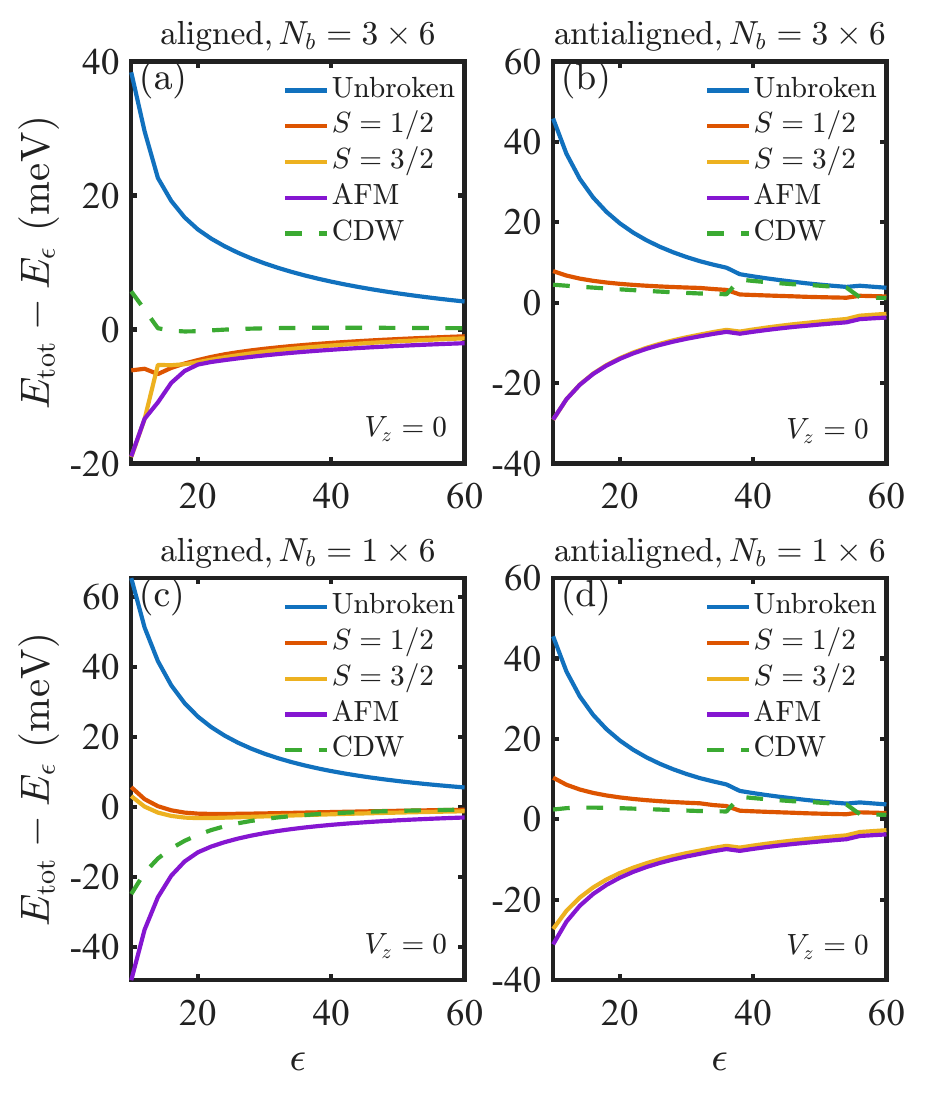}
    \caption{Hartree-Fock comparison of ground state energies as a function of $\epsilon$ at $V_z=0$. The left and right columns correspond to aligned and antialigned cases, respectively, while the top and bottom rows correspond to calculations with $N_b = 3 \times 6$ (including band mixing) and $N_b = 1 \times 6$ (no band mixing). We compare the unbroken state, spin-polarized states with $S=1/2$ and $S=3/2$, the AFM state, and the CDW state.}
   \label{fig:figEtot}
\end{figure}

\begin{figure}
    \centering
    \includegraphics[trim={1cm 0cm 0cm 0cm},clip,width=1\columnwidth]{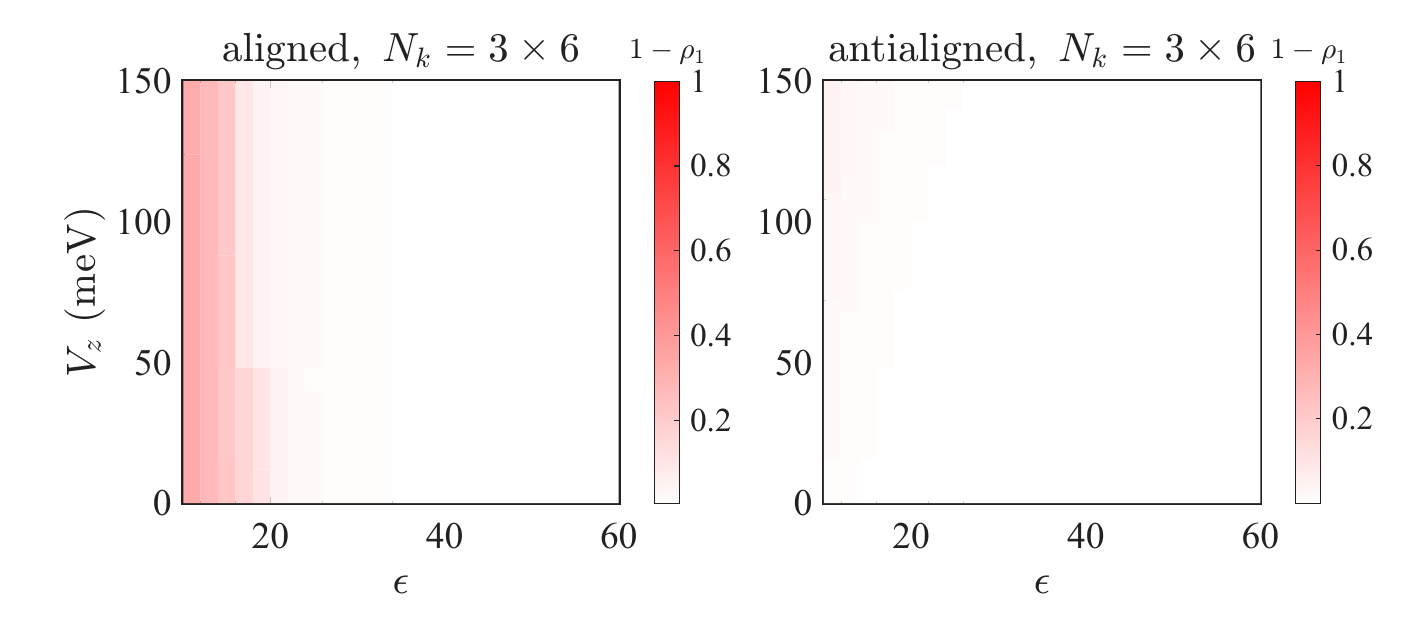}
    \caption{Band mixing parameter $1-\rho_1$ as a function of dielectric constant $\epsilon$ and interlayer bias $V_z$ for (left) aligned and (right) antialigned stacking. The quantity $1-\rho_1$ measures the weight of the wavefunction outside the first moir\'e band in the band basis. Band mixing becomes significant in the aligned case at small $\epsilon$ and large $V_z$, while it remains weak in the antialigned case.}

   \label{fig:band_mixing}
\end{figure}

\section{Additional Hartree-Fock results}

In this section, we provide additional Hartree-Fock results that further illustrate the competition between different broken-symmetry states. The comparison of ground state energies is shown in Fig.~\ref{fig:figEtot}. The left and right columns correspond to the aligned and antialigned stacking configurations, respectively, while the top and bottom rows show results with and without band mixing. In all cases, we include the unbroken state, spin-polarized states with $S=1/2$ and $S=3/2$, as well as antiferromagnetic (AFM) and charge density wave (CDW) states.  

For the aligned case [Figs.~\ref{fig:figEtot}(a) and (c)], AFM order is favored over a broad range of $\epsilon$, while the CDW solution remains close in energy but always higher. Without band mixing [Fig.~\ref{fig:figEtot}(c)], spin-polarized states are not energetically competitive. With band mixing included [Fig.~\ref{fig:figEtot}(a)], however, the energy of the spin-polarized states is significantly lowered, and the ferromagnetic state becomes the ground state at small dielectric constant.  

For the antialigned case [Figs.~\ref{fig:figEtot}(b) and (d)], the overall behavior is qualitatively similar. Without band mixing [Fig.~\ref{fig:figEtot}(d)], the AFM state is energetically favorable. Including band mixing [Fig.~\ref{fig:figEtot}(b)] lowers the energy of the ferromagnetic state, which becomes the ground state at small $\epsilon$, although the influence of band mixing is much weaker compared to the aligned case.

To quantify the role of band mixing, we define the band mixing parameter as the weight of the wavefunction outside the first moir\'e band in the band basis,
\begin{equation}
    1-\rho_1 = 1-\frac{1}{N_m}\sum_{\bm k}\left[\hat{\rho}(\bm k)\right]_{11},
\end{equation}
where the subscript $1$ refers to the first moir\'e band. Figure~\ref{fig:band_mixing} shows the extent of band mixing for both aligned and antialigned cases. We find that band mixing becomes significant in the aligned case at strong Coulomb interaction, while in the antialigned case its effect remains overall weak.

\begin{figure*}[!t]
    \centering
    \includegraphics[trim={0cm 0cm 0cm 0cm},clip,width=2\columnwidth]{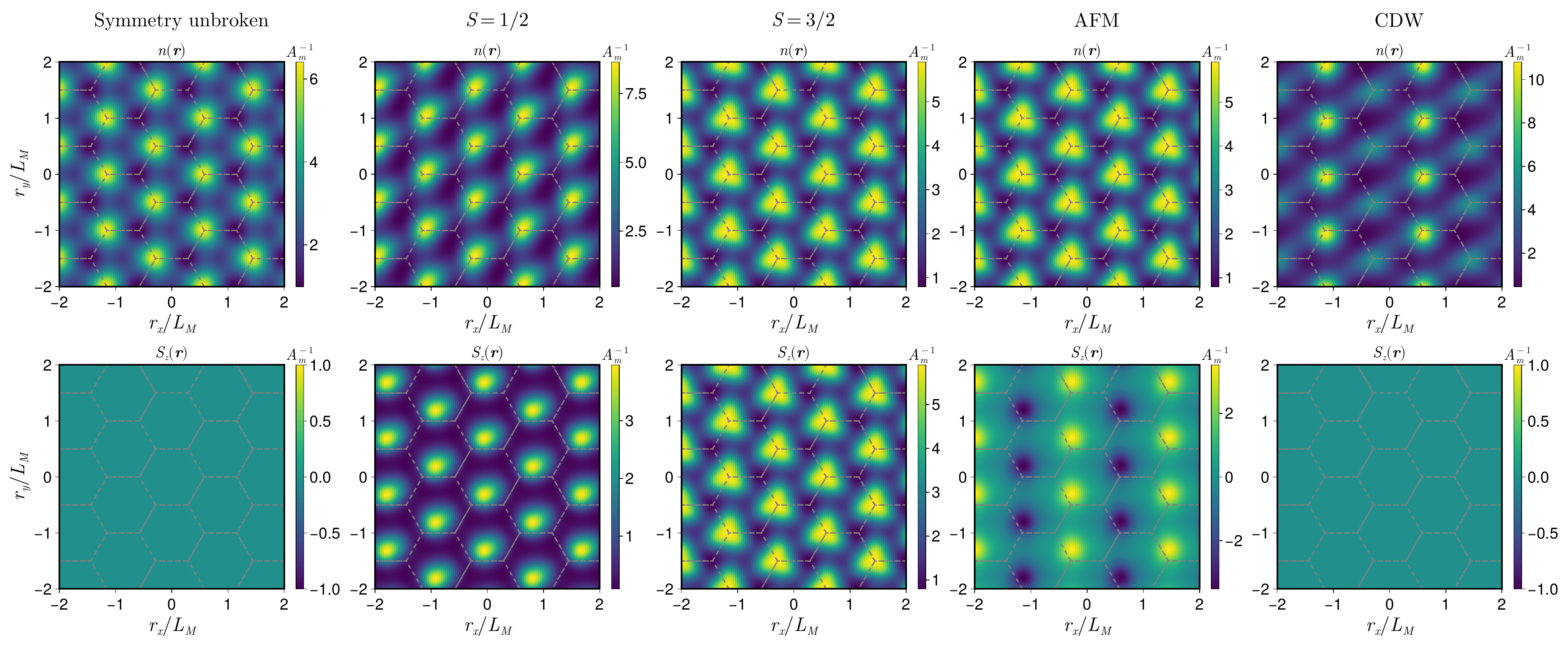}
    \caption{Real-space density and corresponding spin texture for aligned SnSe$_2$ at $V_z = 0$ meV, $\nu=3$, and twist angle $3.89^\circ$. Each column shows the ground state for $S=3/2$, $S=1/2$, and AFM, respectively. The top row shows the real-space density, while the bottom row shows the real-space spin texture. The gray solid lines represent the moir\'e unit cell.}
   \label{fig:figS3}
\end{figure*}

\begin{figure*}[!t]
    \centering
    \includegraphics[trim={0cm 0cm 0cm 0cm},clip,width=2\columnwidth]{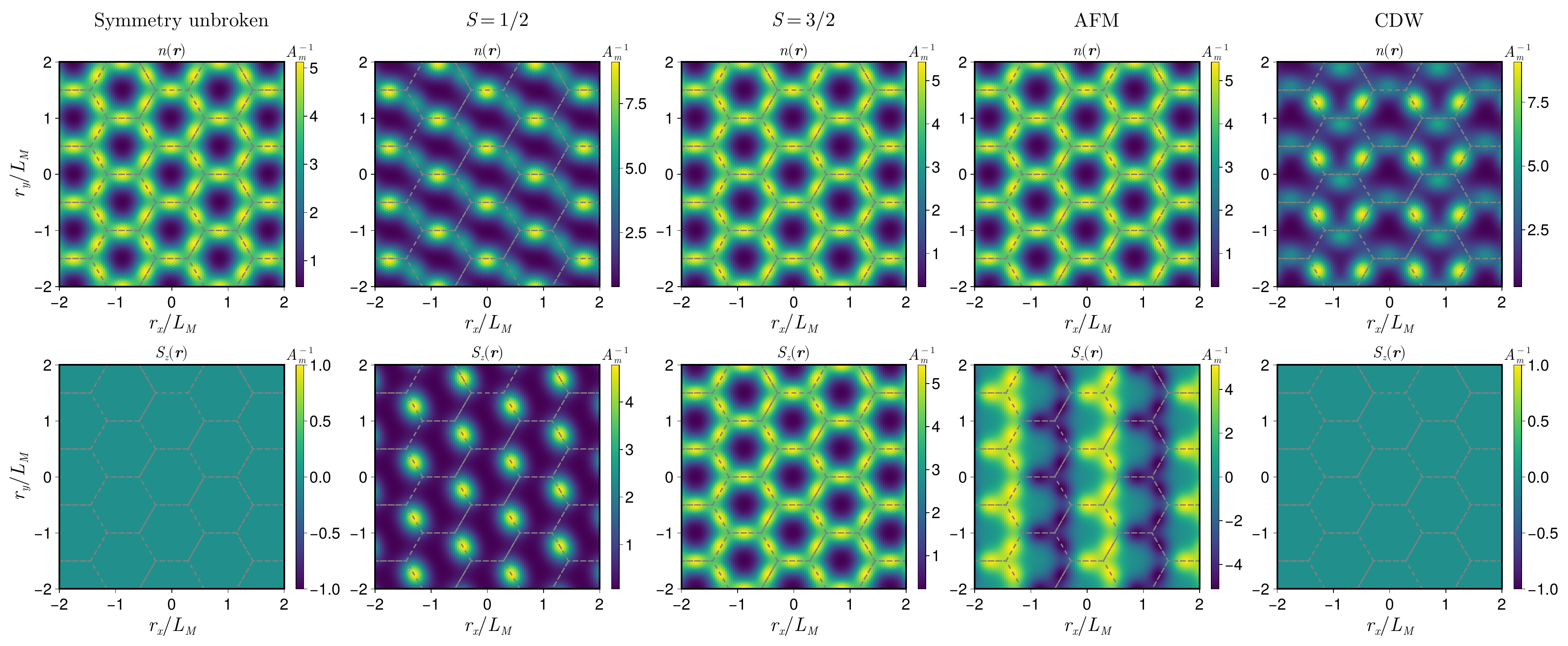}
    \caption{Real-space density and corresponding spin texture for antialigned SnSe$_2$ at $V_z = 0$ meV, $\nu=3$, and twist angle $3.89^\circ$. Each column shows the ground state for $S=3/2$, $S=1/2$, and AFM, respectively. The top row shows the real-space density, while the bottom row shows the real-space spin texture. The gray solid lines represent the moir\'e unit cell.}
   \label{fig:figS4}
\end{figure*}

Finally, to gain further insight into the broken-symmetry solutions, we compute the real-space spin textures. The results are shown in Fig.~\ref{fig:figS3} for the aligned case and Fig.~\ref{fig:figS4} for the antialigned case. Each column corresponds to the lowest-energy broken-symmetry state for symmetry unbroken, $S=3/2$, $S=1/2$, and AFM order, respectively. The top row shows the charge density, while the bottom row displays the real-space spin configuration. These textures illustrate the microscopic differences between the competing ordered states within the moir\'e unit cell.

\end{document}